\documentclass[12pt,a4paper]{article} 

\usepackage{amsfonts}
\usepackage{amstext}
\usepackage{amsmath}
\usepackage{amssymb}
\usepackage{bm,empheq}
\usepackage{amsthm}

\usepackage{tabularx}

\usepackage{graphicx}
\usepackage{color}
\usepackage{here}

\numberwithin{equation}{section}
\numberwithin{figure}{section}
\numberwithin{table}{section}

\newtheorem{thm}{Theorem}[section]

\newtheorem{prop}[thm]{Proposition}

\theoremstyle{definition}

\def\IP{\mathbb {P}}

\newcommand{\rf}[1]{(\ref{#1})}
\newcommand{\beq}{\begin{equation}}
\newcommand{\eeq}{\end{equation}}
\newcommand{\bea}{\begin{eqnarray}}
\newcommand{\eea}{\end{eqnarray}}

\renewcommand{\a}{\alpha}

\newcommand{\calF}{\mathcal{F}{\negdbltinyspace}}

\newcommand{\halftinyspace}{\hspace{0.0278em}} 
\newcommand{\tinyspace}{\hspace{0.0556em}}
\newcommand{\trehalftinyspace}{\hspace{0.0834em}}
\newcommand{\dbltinyspace}{\hspace{0.1112em}}
\newcommand{\trpltinyspace}{\hspace{0.1668em}}
\newcommand{\qdrpltinyspace}{\hspace{0.2224em}}
\newcommand{\qntpltinyspace}{\hspace{0.2780em}}
\newcommand{\neghalftinyspace}{\hspace{-0.0278em}}
\newcommand{\negtinyspace}{\hspace{-0.0556em}}
\newcommand{\negtrehalftinyspace}{\hspace{-0.0834em}}
\newcommand{\negdbltinyspace}{\hspace{-0.1112em}}
\newcommand{\negtrpltinyspace}{\hspace{-0.1668em}}
\newcommand{\negqdrpltinyspace}{\hspace{-0.2224em}}
\newcommand{\negsxpltinyspace}{\hspace{-0.3336em}}

\newcommand{\onethird}{\frac{1}{3}}

\newcommand{\quarter}{\frac{1}{4}}

\newcommand{\define}{\leftdefine}
\newcommand{\leftdefine}{:=}

\newcommand{\bra}[1]{\langle #1 |}

\newcommand{\ket}[1]{| #1 \rangle}

\newcommand{\vac}{\bra{{\rm vac}}}
\newcommand{\cuum}{\ket{{\rm vac}}}
\newcommand{\expect}[1]{\langle #1 \tinyspace\rangle}

\newcommand{\combi}[2]{\left( \!\! \begin{array}{c} 
	\raise0.5ex\hbox{$#1$} \\ \lower0.5ex\hbox{$#2$} \\ 
	\end{array} \!\! \right)}

\newcommand{\commutator}[2]{[\, #1 {\dbltinyspace}, #2 \,]}

\newcommand{\commutatorBig}[2]{\Big[\, #1 \,,\, #2 \,\Big]}

\newcommand{\cc}{\mu} 

\font\twelvemsbm = msbm10 scaled\magstep1
\font\tenmsbm = msbm10
\font\eightmsbm = msbm8
\font\sixmsbm = msbm6
\newcommand{\Dbl}[1]{\leavevmode\raise-.10ex\hbox{\twelvemsbm #1}}
\newcommand{\dbl}[1]{\leavevmode\raise-.00ex\hbox{\tenmsbm #1}}
\newcommand{\dblsmall}[1]{\leavevmode\raise-.05ex\hbox{\eightmsbm #1}}
\newcommand{\dbltiny}[1]{\leavevmode\raise-.05ex\hbox{\sixmsbm #1}}

\newcommand{\pder}[1]{\frac{\partial}{\partial #1}}

\newcommand{\Hop}{H}

\newcount\figX
\newcount\figY
\newcount\figXl
\newcount\figXr
\newcount\figXll
\newcount\figXrr
\newcount\figXlll
\newcount\figXrrr
\newcommand{\minipagewidth}{\linewidth}
\newcommand{\figwidth}{\unitlength}

\newcommand{\pictureY}{141.4}

\newcommand{\figuresize}[3]{
  \figX=#2
  \multiply \figX by -1
  \advance  \figX by 100
  \divide   \figX by 2
  \figY=0
  \renewcommand{\minipagewidth}{#1\linewidth}
  \setlength\unitlength{#1\linewidth}
  \setlength\unitlength{0.01\unitlength}
  \renewcommand{\figwidth}{#2\unitlength}
  \renewcommand{\pictureY}{#3}
}

\newcommand{\figureshift}[2]{
  \figXl=#1
  \advance  \figXl by \figX
  \figXr=#1
  \multiply \figXr by -1
  \advance  \figXr by \figX
  \figXll=#1
  \multiply \figXll by 2
  \advance  \figXll by \figX
  \figXrr=#1
  \multiply \figXrr by -2
  \advance  \figXrr by \figX
  \figXlll=#1
  \multiply \figXlll by 3
  \advance  \figXlll by \figX
  \figXrrr=#1
  \multiply \figXrrr by -3
  \advance  \figXrrr by \figX
  \advance  \figY by #2
}

\newcommand{\T}{T} 

\newcommand{\GG}{{\cal G}}
\newcommand{\NN}{N}

\newcommand{\E}{{\rm e}}

\def\theequation{\arabic{section}.\arabic{equation}}
\def\thefigure{\arabic{figure}}

\allowdisplaybreaks

\begin{document}
\topmargin 0pt
\oddsidemargin 5mm
\headheight 0pt
\headsep 0pt
\topskip 9mm



\begin{center}
  \vspace{24pt}
  {\large \bf
	Multicritical Dynamical Triangulations \\
	and Topological Recursion
    }

  \vspace{24pt}

  {\sl Hiroyuki Fuji}

  \vspace{6pt}

{\small
  Center for Mathematical and Data Sciences and Department of Mathematics\\
   Kobe University\\
   Rokko, Kobe 657-8501, Japan
}
\vspace{6pt}

  \vspace{12pt}

  {\sl Masahide Manabe}

  \vspace{6pt}

{\small
  Liberal Arts \& Data Science Unit\\
  Tottori University\\
  4-101 Koyama-cho Minami, Tottori, 680-8550, Japan
}

  \vspace{12pt}

  and

  \vspace{12pt}

  {\sl Yoshiyuki Watabiki}

  \vspace{6pt}

{\small
  Department of Physics\\
  Institute of Science Tokyo\\
  Oh-okayama 2-12-1, Meguro-ku, Tokyo 152-8551, Japan
}

\end{center}
\vspace{24pt}

\vfill

\begin{center}
  {\bf Abstract}
\end{center}

\vspace{12pt}

\noindent
We explore a continuum theory of multicritical dynamical triangulations and causal dynamical triangulations in two-dimensional quantum gravity from the perspective of the Chekhov-Eynard-Orantin topological recursion.
In the mode-expanded formulation, the Hamiltonian of the former is expressed in terms of a specific $W_{-2}^{(3)}$ operator in a two-reduced $W^{(3)}$ representation, whereas that of the latter is expressed in terms of the corresponding $W_{-2}^{(3)}$ operator in the full $W^{(3)}$ representation.
Starting directly from the string-field-theoretic Hamiltonian formalism, we derive the Schwinger-Dyson equations for both models.
The Schwinger-Dyson equations admit one-cut disk solutions compatible with the physical continuum branches known from the corresponding matrix-model descriptions, and we show that the perturbative amplitudes around these solutions are generated by the Chekhov-Eynard-Orantin topological recursion with explicit spectral curves.

\vfill

\thispagestyle{empty}

\newpage

\setcounter{page}{1}

\tableofcontents

\newpage

\section{Introduction}

The multicritical one-matrix model was developed as an extension of the one-matrix
model describing pure gravity with central charge $c \!=\! 0$ \cite{MM:Kaza}. Unlike the pure-gravity
matrix model, which consists solely of triangles, the multicritical model incorporates
not only triangles but also higher polygons, such as quadrilaterals and pentagons, in
order to realize higher-order critical points. By taking the continuum limit around such
a critical point, one obtains a continuum theory of two-dimensional quantum gravity
related to the $(2,2m-1)$ minimal series. This model was later reformulated as a string
field theory \cite{SFT:GK,SFT:Watabiki}, and, after a mode expansion, its Hamiltonian was
expressed in terms of operators associated with a $W^{(3)}$ representation \cite{SFT:AW}.
In this paper, we begin our analysis from this mode-expanded Hamiltonian formulation.

Pure Dynamical Triangulations (DT) consists of triangulations of a two-dimensional
surface with equilateral triangles of fixed size. It should be noted that not all versions
of pure DT are equivalent to matrix models; for particular choices of discrete
configurations, one can establish an exact correspondence between such pure DT ensembles
and specific matrix models. In the continuum limit, all minor variants of pure DT
coincide with the corresponding matrix model, and moreover, both are believed to
coincide with Liouville gravity without matter fields, although a rigorous proof of this
coincidence has not yet been established. At the same time, pure DT also admits a
continuum Hamiltonian formulation constructed directly from the dynamical-triangulation
framework, without using matrix models as an intermediate step. For pure DT, it has
long been known that after the mode expansion, the continuum Hamiltonian can be
expressed in terms of a specific $W_{-2}^{(3)}$ mode in a two-reduced $W^{(3)}$
representation \cite{MM:FKN,MM:DVV,SFT:AW}, and this observation was also employed
in our previous study \cite{FMW2025a}. In the present paper, pure DT serves as the basic
example from which we proceed to its multicritical generalization.

We refer to the continuum theory associated with the $(2,2m-1)$ minimal string as the
$m$-th multicritical DT. The first aim of this paper is to investigate this theory from the
viewpoint of the Chekhov-Eynard-Orantin (CEO) topological recursion
\cite{Eynard:2004mh,Chekhov:2006vd,Eynard:2007kz}, starting directly from the
string-field-theoretic Hamiltonian formulation.

However, this class of theories suffers from the absence of a genuine time direction\footnote{%
Although in the proper gauge the geodesic distance
may play a role similar to that of a time coordinate,
it does not represent a genuine causal time.
In other words, the notion of causality, which is essential for describing physical spacetime,
cannot be properly implemented within the DT framework.}
and therefore does not provide an adequate description of spacetime dynamics. To address
this issue, Causal Dynamical Triangulations (CDT) was introduced as a model incorporating
an explicit time direction \cite{CDT:AL}. At the discrete level, CDT may be viewed as a restricted
version of DT in which additional constraints are imposed so as to distinguish space-like
and time-like links, while configurations for which such a distinction cannot be made
consistently are excluded.

At the early stage of its development, CDT was too simple in structure to become a major
subject of investigation, because string interactions were entirely absent. Consequently,
two directions of extension were pursued. One was its extension to three- and
four-dimensional spacetimes, while the other was a generalization incorporating
interactions among one-dimensional universes. The former turned out to be so difficult
that analytical methods were essentially inapplicable, and the corresponding studies had
to rely mainly on numerical simulations; this line of research later developed into what
is now called higher-dimensional CDT. The latter introduced into CDT the interactions
among one-dimensional universes that had already been present from the outset in DT,
and this extended model is called Generalized CDT (GCDT) \cite{CDT:SFT:ALWWZ}. Furthermore, in the
continuum limit, this model has been shown to be equivalent to a one-matrix model,
although the critical point at which the continuum limit is taken differs from that of DT
\cite{CDT:MM:ALWWZ,CDT:MM:ABW}.

Motivated by the mode-expanded Hamiltonian formulation of the string field theory for DT,
in which the Hamiltonian is expressed through a specific $W_{-2}^{(3)}$ operator in the
two-reduced $W^{(3)}$ representation, we carry out an analogous expansion for GCDT. In
contrast to the DT case, the corresponding Hamiltonian in GCDT involves the
$W_{-2}^{(3)}$ operator defined in the full $W^{(3)}$ representation. From the viewpoint of
this mode-expanded operator formulation, this pattern is opposite in direction to the
familiar inclusion relation in the KP hierarchy: in the standard integrable-hierarchy setting
one has, for example (see, e.g., \cite{JM83}),
\begin{equation}
\mbox{KP} \supset \mbox{KdV}
\qquad
\mathrm{and\;hence}
\qquad
W_{\rm full} \supset W_{\textrm{two-reduced}}
\,.
\nonumber
\end{equation}
In the present dynamical-triangulation framework, by contrast, the DT Hamiltonian is
expressed through the relevant $W_{-2}^{(3)}$ operator in the two-reduced representation,
whereas the corresponding CDT (or GCDT) Hamiltonian is expressed through the operator
in the full representation.%
\footnote{
Here $W^{(3)}$ refers to the operator representation arising from the mode expansion of the
string-field-theoretic Hamiltonian for dynamical triangulations.
In the derivations below, we use the explicit Hamiltonians, whose mode-expanded forms
involve the relevant $W_{-2}^{(3)}$ operators, together with the oscillator commutation
relations; the commutation relations of the $W^{(3)}$ algebra themselves are not invoked
directly. These operators should be distinguished from the cut-and-join type $W$-operators
acting on time variables or generating functions, as well as from the $W$-algebraic
structures associated with quantum Airy structures. At present, the direct relation between
these structures and the $W^{(3)}$ operators considered here is not clear.
}
Thus, in the present mode-expanded formulation, the distinction between DT and CDT is
reflected in whether the relevant Hamiltonian operator is realized in the two-reduced or the
full $W^{(3)}$ representation.

\begin{figure}[t]
\begin{center}
\hspace*{2.0cm}
\includegraphics[width=17cm,pagebox=cropbox,clip]{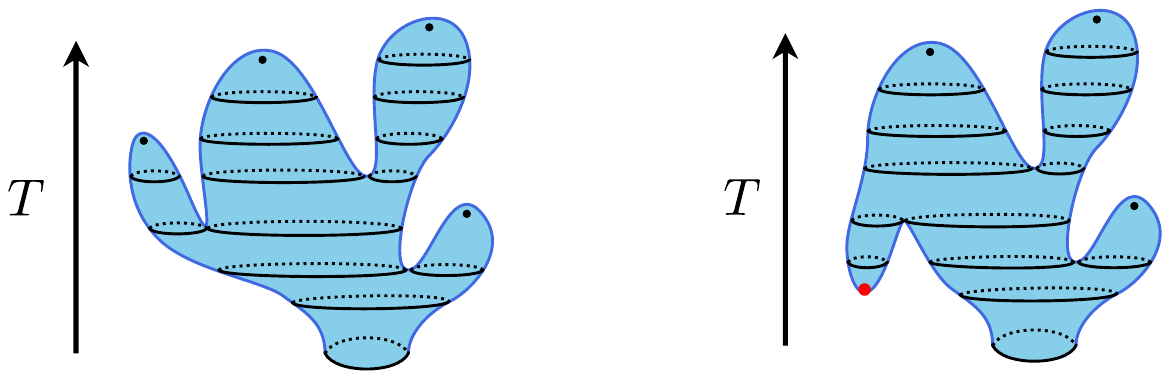}
\vspace{-9cm}
  \caption{Two configurations of a 2D surface with a boundary. \\
$\bullet$ Left panel: An example of a DT or CDT configuration
that satisfies the no-big-bang condition.
(The configuration does not create a universe from nothing; configurations of the type shown in the right panel are therefore excluded.) \\
$\bullet$ Right panel: An example of a CDT configuration
that does not satisfy the no-big-bang condition.
(The red point represents the creation of a universe from nothing.
The region including the red point and its surroundings
is inaccessible from the lower-right universe,
indicating that this configuration preserves causality.)
}
  \label{fig:DTpicture}
\end{center}
\end{figure}

However, unlike DT, an ambiguity already arises at the level of the definition of CDT.
To resolve this ambiguity, two possible conditions have been proposed. One is to define
the Hamiltonian so that it satisfies the no-big-bang condition, which forbids the creation
of universes from nothing (see the left panel of Fig.\ \ref{fig:DTpicture}), and the other is to identify the
Hamiltonian directly with the $W^{(3)}$ operator \cite{CDT:SFT:AW}.

In the present paper, we adopt the no-big-bang condition and focus on the
two-dimensional GCDT model, which, for simplicity, we will refer to as CDT. We
consider its multicritical extension \cite{Atkin:2012ka}, which we call the $m$-th
multicritical CDT, as the causal analogue of the $m$-th multicritical DT. This model also
admits a matrix-model description and provides an extended framework that includes the
models in \cite{CDT:SFT:ALWWZ,CDT:AGGS} as special cases.%
\footnote{
For $m=2$, the $m$-th multicritical DT (resp. CDT) reduces to the pure DT (resp. CDT) in the continuum limit.}

The second aim of this paper is to derive the corresponding Schwinger-Dyson (SD) equations directly from the Hamiltonian formalism and to show that they admit a one-cut disk solution whose perturbative amplitudes are generated by the CEO topological recursion.
The CEO topological recursion is rooted in matrix models and was formulated in
\cite{Eynard:2007kz} purely in terms of spectral-curve data $(\Sigma;\xi,y,B)$, where
$\Sigma$ is a compact Riemann surface, $\xi$ and $y$ are meromorphic functions on
$\Sigma$, and $B$ is a meromorphic symmetric bidifferential on $\Sigma^2$. Here the zeros
of $d\xi$ are assumed to be simple and different from the zeros of $dy$. From such
spectral-curve data, the CEO topological recursion defines symmetric multidifferentials
$\omega^{(h)}_N$ on $\Sigma^N$, labeled by two integers $h\ge0$ and $N\ge1$ satisfying
$2h-2+N>0$. The spectral-curve data relevant to the $m$-th multicritical DT and CDT
models discussed in this paper are as follows:
$$
\Sigma = \mathbb{P}^1\,,
\quad
B(\eta_1, \eta_2) = \frac{d\eta_1d\eta_2}{\left(\eta_1-\eta_2\right)^2}\,,
\quad \eta_1, \eta_2 \in \mathbb{P}^1\,,
$$
and
\begin{align*}
m\textrm{-th multicritical DT}:\quad&
\xi=\xi(\eta)
=\eta^2 - \alpha\,,
\quad \eta \in \mathbb{P}^1\,,
\\
&
y=
\left(\xi(\eta)^{m-1}+\sum_{p=1}^{m-1}t_{m-p}\, \xi(\eta)^{p-1} \right)
\eta
\,,
\\
m\textrm{-th multicritical CDT}:\quad&
\xi=\xi(\eta)
=-\frac{\alpha+\beta}{2}
+\frac{\alpha-\beta}{4}\left(\eta+\eta^{-1}\right),
\quad \eta \in \mathbb{P}^1\,,
\\
&
y=
\frac{1}{2g}\left(\xi(\eta)^{m-1}+\sum_{p=1}^{m-1}t_{m-p}\, \xi(\eta)^{p-1} \right)
\frac{\alpha-\beta}{4}\left(\eta-\eta^{-1}\right).
\end{align*}
Here $\alpha=\alpha(\mu)$, $t_1=t_1(\mu),\ldots,t_{m-1}=t_{m-1}(\mu)$ in the
$m$-th multicritical DT depend on $\mu=\{\mu_2,\ldots,\mu_m\}$, which denotes the
cosmological constants, while $\alpha=\alpha(\mu)$, $\beta=\beta(\mu)$,
$t_1=t_1(\mu),\ldots,t_{m-1}=t_{m-1}(\mu)$ in the $m$-th multicritical CDT are
functions of $\mu=\{\mu_2,\ldots,\mu_m\}$ and $g$.%
\footnote{The parameter $g$ in CDT is introduced to count the number of branches of a 2D surface.}
The multidifferential $\omega^{(h)}_N(\xi_1,\ldots,\xi_N)=\omega^{(h)}_N(\xi(\eta_1),\ldots,\xi(\eta_N))$ for each model
then provides the amplitude associated with a two-dimensional surface with $h$ handles
and $N$ boundaries.%
\footnote{
In \cite{FMW3}, we propose a generalization of the Hamiltonian formalism of two-dimensional quantum gravity, which is discussed in \cite{FMW2025a} and in the present paper, using the CEO topological recursion as a guiding principle.
}
On the DT side, we will later show that, for a distinguished specialization of the
cosmological constants, this spectral-curve description also reproduces the
conformal-background amplitudes of the $(2,2m-1)$ minimal string.

From this viewpoint, the appearance of a topological-recursive structure is not unexpected.
The correspondence between the SD equations in the string-field-theory formulation and
the matrix-model loop equations is already natural from the peeling/transfer-matrix
construction of dynamical triangulations \cite{SFT:Watabiki} and from the correspondence
between the loop creation operator and the matrix-model resolvent exhibited in our
previous work \cite[Appendix D]{FMW2025a}. More generally, the relation between
matrix-model loop equations and topological recursion is well established
\cite{Eynard:2004mh,Chekhov:2006vd,Eynard:2007kz}.
These observations motivate us to ask whether the SD equations obtained directly from the multicritical DT and CDT Hamiltonians admit one-cut disk solutions whose perturbative amplitudes are generated by the CEO topological recursion with explicit spectral curves, without passing through a matrix-model realization.

To place this question in context, it is important to distinguish carefully between what
had already been established in the previous literature and what is proved in the present
paper. The identification of multicritical one-matrix models with Liouville gravity coupled
to the $(2,2m-1)$ minimal conformal models was established through continuum analyses
of matrix models and their string equations \cite{MM:Kaza,MM:Stau}. A continuum
string-field-theoretic description of DT was subsequently proposed in
\cite{Ishibashi:1993nq}, where a Hamiltonian formalism describing the splitting and
joining of loops was introduced and the associated SD equations reproducing the
continuum loop equations of matrix models were derived. It was later pointed out in
\cite{SFT:Watabiki} that the original Hamiltonian is not well defined without
regularization, and a regularized continuum Hamiltonian was constructed directly from
the dynamical-triangulation framework. Building on this formulation, the mode-expanded
Hamiltonian framework was further extended to multicritical DTs associated with the
$(2,2m-1)$ series \cite{SFT:AW}.

These developments established the relevant Hamiltonian framework, but the direct
relation to the amplitudes of the corresponding minimal string, together with an explicit
form of the Hamiltonian-derived SD equations suitable for topological recursion, had not
been made equally explicit. For pure CDT, a string-field-theoretic Hamiltonian formalism
was constructed in later works \cite{CDT:SFT:ALWWZ}, and the resulting SD equations
were shown to agree with those of the corresponding matrix models
\cite{CDT:MM:ALWWZ,CDT:MM:ABW}.
What remained to be clarified was whether, for the multicritical DT and CDT Hamiltonians considered here, the SD equations derived directly from the Hamiltonian formalism admit one-cut disk solutions and whether the perturbative amplitudes around these solutions are generated by the CEO topological recursion without recourse to matrix models.

The novelty of the present work is therefore not a new general equivalence between SD
equations and topological recursion, but the direct realization of this structure for the
specific multicritical DT and CDT Hamiltonians considered here. More concretely, we
formulate the $m$-th multicritical DT and, as its causal analogue, the $m$-th multicritical
CDT under the no-big-bang condition within a common mode-expanded Hamiltonian
formalism, and derive their SD equations directly from the Hamiltonian formalism.
We show that these SD equations admit one-cut disk solutions and that the perturbative amplitudes around these solutions are generated by the CEO topological recursion with explicit spectral curves.
The two models are distinguished by the two-reduced and full $W^{(3)}$
operator realizations described above, and our multicritical CDT model includes, as special
cases, previously studied models such as those discussed in
\cite{Atkin:2012ka,CDT:AGGS}.

On the DT side, we further clarify how the conformal background of the Hamiltonian
formulation is related to the $(2,2m-1)$ minimal string. In particular, we identify a
distinguished specialization of the cosmological constants for which the disk amplitude
computed from the Hamiltonian formalism coincides exactly with that of the $(2,2m-1)$
minimal string in the conformal background. The corresponding specialized spectral curve
agrees, up to the conventions and normalizations used here, with the standard
conformal-background minimal-string spectral curve known from previous approaches
\cite{Gregori:2021tvs,Seiberg:2003nm}. What is specific to the present analysis is its
direct emergence from the string-field-theoretic Hamiltonian together with the explicit
identification of the Hamiltonian cosmological constants required to reach this conformal
background.

We emphasize the logical status of the ingredients used in this construction. Once the
Hamiltonian is fixed, the formal large-$T$ stationarity prescription together with the
oscillator commutation relations leads to the SD equations, where $T$ denotes the time
variable in the Hamiltonian formalism. The subsequent expansion in $\GG$ is a formal
genus expansion, consistent with the role of $\GG$ as the parameter counting handles.
The Hamiltonian-derived disk SD equations admit one-cut solutions of the form used below.

In the DT case, we focus on the one-cut solution because its analytic
structure is consistent with that of the disk amplitude obtained in the
continuum limit of multicritical one-matrix models.
In the CDT case, we regard the
multicritical one-cut solution as the natural analogue of the physical
continuum branch selected in pure CDT
\cite{MM:Kaza,CDT:MM:ALWWZ,Atkin:2012ka,CDT:AGGS}.
We do not claim that the SD equations uniquely determine the analytic
structure of the disk amplitude. For the one-cut solutions considered
here, the corresponding data are fixed by the Hamiltonian couplings
through the algebraic relations derived below, while the asymptotic and
analyticity conditions identify the physical solutions. In the CDT case,
the no-big-bang condition specifies the Hamiltonian, whereas the no-pole
condition selects the physical branch among the resulting algebraic
solutions.

Together with the conformal-background identification above, these results provide a
nontrivial amplitude-level consistency check of the $m$-th multicritical DT Hamiltonian.
They also show that the higher perturbative amplitudes obtained from the corresponding
CEO topological recursion admit a direct interpretation as amplitudes of quantum gravity
within the string-field-theoretic formalism. As a further consequence, after an appropriate
normalization, the large-$m$ limit of this conformal-background amplitude reproduces the
JT-gravity disk amplitude after Laplace transformation. Since the corresponding
spectral-curve data coincide with those of JT gravity in this limit, the associated
perturbative topological recursion becomes equivalent to that of JT gravity itself.

This paper is organized as follows. In Section~\ref{sec:DT}, after reviewing pure DT and
its mode-expanded Hamiltonian formulation, we develop the $m$-th multicritical DT
analysis, the resulting one-cut disk solution, and its CEO-recursive perturbative expansion.
In Section~\ref{sec:MultiDTandJTgravity}, we identify a distinguished specialization of the
cosmological constants for which the disk amplitude reproduces the conformal-background
amplitude of the $(2,2m-1)$ minimal string, and we show that its large-$m$ limit yields the
Laplace dual of the JT-gravity disk amplitude after an appropriate normalization. In
Section~\ref{sec:CDT}, after reviewing the mode-expanded Hamiltonian formulation of pure
CDT, we develop the corresponding multicritical CDT analysis. In Appendices
\ref{app:proof_prop} and \ref{app:proof_prop_cdt}, we present proofs of several
propositions, and in Appendix~\ref{app:list_amp} we list several amplitudes computed by
the CEO topological recursion.

\section{Dynamical Triangulation (DT)}
\label{sec:DT}

In this section, we introduce the $m$-th multicritical DT whose Hamiltonian is expressed through a specific $W_{-2}^{(3)}$ operator in the two-reduced $W^{(3)}$ representation, and show that the Hamiltonian-derived SD equations admit a one-cut disk solution whose perturbative amplitudes in the parameter $\GG$, which counts the number of handles of a 2D surface, are generated by the CEO topological recursion.
We first review the case $m=2$, which corresponds to pure DT.

\subsection{Pure DT}
\label{sec:PureDT}

\subsubsection{Background}\label{sec:PureDT_bg}

Before reviewing the formulation of pure DT, let us briefly recall the theoretical background relevant to the SD equations in this case. 
The relation between pure DT, matrix models, and Liouville gravity has long been studied in the continuum limit. 
In parallel with this line of development, however, the continuum Hamiltonian
formalism of pure DT was obtained directly as the continuum limit of a discrete model of
two-dimensional surfaces built from triangulations \cite{SFT:Watabiki}.
In this formulation, the underlying discrete model is not itself given in an explicit
matrix-model form, and the continuum Hamiltonian was constructed within the
string-field-theoretic approach without using matrix models as an intermediate
step.

Moreover, the mode expansion of the string field shows that the continuum Hamiltonian of pure DT can be expressed in terms of the specific $W_{-2}^{(3)}$ operator in the two-reduced $W^{(3)}$ representation \cite{SFT:AW}. Thus, in the pure case, the relation between the Hamiltonian description and the resulting SD equations is already explicit within the dynamical-triangulation formalism through this mode-expanded operator representation.

In this section, we summarize the Hamiltonian and operator formalism of pure DT, review the amplitudes and SD equations that follow from it, and fix the notation and conventions used in the later sections. As shown in our previous work \cite{FMW2025a}, these SD equations admit perturbative solutions given by the CEO topological recursion.

\subsubsection{Formulation}
\label{sec:PureDTdef}

In \cite{SFT:AW}, a continuum theory of pure DT was formulated.
Here we briefly summarize this formulation.

In the continuum limit, 
the generating function of amplitudes for pure DT is given by
\begin{equation}\label{GeneratingFunModeExpansion}
Z[j]  \ \define \
\lim_{\T\rightarrow\infty}
  \vac {\tinyspace} \E^{-T \Hop}
    \exp{\negtrpltinyspace}\bigg(
      \sum_{\ell=1}^\infty \phi^\dagger_\ell j_\ell
    {\negtinyspace}\bigg)
  {\negtinyspace}\cuum
\,,
\end{equation}
where 
$\phi^\dagger_\ell$ and $\phi_\ell$
[\,$\ell {\negdbltinyspace}\in{\negdbltinyspace} \Dbl{N}$\,]
are quantum operators that obey the commutation relations:
\begin{equation}\label{CommutationRelationPhi}
\commutator{\phi_k}{\phi^\dagger_n}
= \delta_{{\tinyspace}k,n}
\,,
\qquad
\commutator{\phi_k}{\phi_n}
=
\commutator{\phi^\dagger_k}{\phi^\dagger_n}
=
0
\,,
\qquad
\mbox{[\,$k$, $n {\negdbltinyspace}\in{\negdbltinyspace} \Dbl{N}$\,]}
\,,
\end{equation}
and 
the vacuum state
$\vac$ and its dual $\cuum$ 
satisfy the conditions: 
\begin{equation}\label{vacuumCondition_phi}
\expect{{\rm vac}|{\rm vac}} = 1
\,,
\qquad
\vac \phi^\dagger_\ell = 0
\,,
\qquad
\phi_\ell \cuum = 0
\,,
\qquad
\mbox{[\,$\ell {\negdbltinyspace}\in{\negdbltinyspace} \Dbl{N}$\,]}
\,.
\end{equation}
Here  
$\T$ denotes the proper time, 
and
$\Hop$ denotes the Hamiltonian, defined by
\begin{eqnarray}\label{PureDT_HamiltonianModeExpansion}
\Hop \!\!&\define&\!\!
   -{\qntpltinyspace}
       \frac{\GG}{4}{\tinyspace} \phi_{{\halftinyspace}4}
 \,-\, \Big(
         \cc_2
         -
         \frac{\GG}{2}{\dbltinyspace}\phi_1
       \Big)^{{\negtrpltinyspace}2}
       \phi_{{\tinyspace}2}
 -\> \sum_{\ell=1}^\infty \phi_{\ell+1}^\dagger {\tinyspace} \ell \phi_\ell
 \,+\, \cc_2
     \sum_{\ell=4}^\infty \phi_{\ell-3}^\dagger {\tinyspace} \ell \phi_\ell
\nonumber\\&&\!\!
 -\> \frac{1}{2}
  \sum_{\ell=6}^\infty {\tinyspace} \sum_{n=1}^{\ell-5}
  \phi_n^\dagger \phi_{\ell-n-4}^\dagger {\tinyspace}
  \ell \phi_\ell
\,-\,
  \frac{\GG}{4}
  \sum_{\ell=1}^\infty {\tinyspace} \sum_{n=\max(5-\ell,1)}^\infty\!\!
  \phi_{n+\ell-4}^\dagger {\tinyspace}
  n {\halftinyspace} \phi_n {\tinyspace} \ell \phi_\ell
\,,
\end{eqnarray}
where 
$\cc_2 {\negdbltinyspace}\define{\negdbltinyspace} \frac{3}{8} \cc$, 
$\cc$ denotes the cosmological constant, and the parameter
$\GG$ is introduced to count the number of handles of a 2D surface. 
The Hamiltonian \rf{PureDT_HamiltonianModeExpansion} 
satisfies the so-called no-big-bang condition, 
\begin{equation}\label{NoBigBangCondition}
\Hop {\dbltinyspace} \cuum  \,=\,  0
\,,
\end{equation}
which implies the stability of the vacuum $\cuum$. 
The variable $T$, which is conjugate to the Hamiltonian (\ref{PureDT_HamiltonianModeExpansion}) in pure DT, is regarded as the geodesic distance from the entrance loop, where the entrance loop is depicted at the bottom of the left panel of Fig.\ \ref{fig:DTpicture}. 
Consequently, the variable $T$ lacks a causal interpretation, because every point on the two-dimensional surface is accessible from the entrance loop by the Hamiltonian evolution.
This property, which forbids the creation of any one-dimensional universe from nothing, is implemented in the Hamiltonian formulation by the no-big-bang condition.

Throughout this paper, the $T\to\infty$ limit in \eqref{GeneratingFunModeExpansion}, as well as in the amplitude definitions below, is understood formally as selecting the stationary large-$T$ sector of the continuum evolution, rather than as an analytically established convergent limit of the finite-$T$ expressions.
In particular, we do not claim a rigorous analytic proof of the existence or finiteness of these limits. The SD equations and the subsequent perturbative expansion in $\GG$ are understood in this formal sense.

We now introduce the star operation $\star$ by
\begin{equation}
\star:  \mathbb{C}[\phi,\phi^{\dagger}] \longrightarrow \mathbb{C}[j,\partial/\partial j]\,,
\end{equation}
where $\phi=\{\phi_n\}$, $\phi^{\dagger}=\{\phi_n^{\dagger}\}$ and $j=\{j_n\}$, $\partial/\partial j=\{\partial/\partial j_n\}$.
The star operation for the operators $\phi_{\ell}^{\dagger}, \phi_{\ell}$ is given by
\begin{equation}\label{StarOpPsiModeExpansion}
  \bigl( \phi^\dagger_\ell \bigr)^\star \,=\, \pder{j_\ell}
\,,
\qquad
  \bigl( \phi_\ell \bigr)^\star \,=\, j_\ell
\,,
\qquad
\mbox{[\,$\ell {\negdbltinyspace}\in{\negdbltinyspace} \Dbl{N}$\,]}
\,.
\end{equation}
The action of the star operation on the generating function of amplitudes (\ref{GeneratingFunModeExpansion}) is given by
\begin{equation}\label{StarOpDefModeExpansion}
A_N^\star \ldots {\tinyspace} 
A_2^\star {\tinyspace} 
A_1^\star {\tinyspace} Z[j]
\,=\,  
\lim_{\T\rightarrow\infty}
  \vac {\tinyspace} \E^{-T \Hop} 
    A_1 A_2 \ldots A_N
    \exp{\negtrpltinyspace}\bigg(
      \sum_{\ell=1}^\infty \phi^\dagger_\ell j_\ell
    {\negtinyspace}\bigg)
  {\negtinyspace}\cuum
\,,
\end{equation}
where $A_1,\ldots, A_N\in \mathbb{C}[\phi,\phi^{\dagger}]$ and $A_1^{\star},\ldots, A_N^{\star} \in \mathbb{C}[j,\partial/\partial j]$.

Here we introduce the operators $\a_n$, defined by{\dbltinyspace}%
\footnote{%
A positive constant
$p {\negdbltinyspace}>{\negdbltinyspace} 0$ 
is introduced for later convenience. 
}
\begin{equation}\label{DefAlphaOperatorPureDT}
\a_n \,\define\,
\left\{
\begin{array}{cl}
\displaystyle
\sqrt{\frac{p}{\GG}} {\dbltinyspace} \pder{j_n}
& \hbox{[\,$n \!>\! 0$\,]}
\rule[-0pt]{0pt}{0pt}\\
\displaystyle
\nu
& \hbox{[\,$n \!=\! 0$\,]}
\rule[-8pt]{0pt}{24pt}\\
\displaystyle
-{\tinyspace} n {\tinyspace}
 \bigg(
   \lambda_{-n} + \sqrt{\frac{\GG}{p}} {\trpltinyspace} j_{-n}
 \bigg)
& \hbox{[\,$n \!<\! 0$\,]}
\end{array}
\right.
\,,
\qquad
\mbox{[\,$n {\negdbltinyspace}\in{\negdbltinyspace} \Dbl{Z}$\,]}
\end{equation}
where $p {\negdbltinyspace}={\negdbltinyspace} 2$ and 
\begin{equation}\label{CoherentEigenValuesPureDT}
\nu \define 0
\,,
\quad
\lambda_1 \define -{\tinyspace}\cc_2 \sqrt{\frac{2}{\GG}}
\,,
\quad
\lambda_{{\tinyspace}5} \define \frac{1}{5}\sqrt{\frac{2}{\GG}}
\,,
\quad
\lambda_{{\tinyspace}n} \define 0
\,,
\quad\mbox{[\,$n {\negdbltinyspace}\in{\negdbltinyspace} \Dbl{N}$, 
              $n {\negtrpltinyspace}\neq{\negtrpltinyspace} 1$,\,$5$\,]}
\,.
\end{equation}
Since the parameters $\lambda_n$ induce a shift $\sqrt{\GG/p} j_{-n}$ in (\ref{DefAlphaOperatorPureDT}), 
the $\lambda_n$ may be regarded as the expectation values of 
$\sqrt{\GG/2}{\trpltinyspace} \phi_n$
[\,$n {\negdbltinyspace}\in{\negdbltinyspace} \Dbl{N}$\,]. 
The operators $\a_n$ satisfy the commutation relation
\begin{equation}\label{CommutationRelationAlphaFlavor}
\commutator{\a_k}{\a_n}
\,=\,
k {\tinyspace}\delta_{{\tinyspace}k+n,0}
\,,
\qquad
\mbox{[\,$k$, $n {\negdbltinyspace}\in{\negdbltinyspace} \Dbl{Z}$\,]}
\,.
\end{equation}

We denote the inverse star operation by $*:\mathbb{C}[j,\partial/\partial j]\to
\mathbb{C}[\phi,\phi^{\dagger}] $, defined such that
$(A^\star)^* {\negtrpltinyspace}={\negtrpltinyspace} A$ for any $A\in\mathbb{C}[\phi,\phi^{\dagger}]$.
The action of the inverse star operation on the differential operators is
\begin{equation}\label{Inv_StarOpPsiModeExpansion}
\left(\pder{j_\ell}\right)^*  \,=\,  \phi^\dagger_\ell 
\,,
\qquad
\left(j_\ell \right)^* \,=\, \phi_\ell  
\,,
\qquad
\mbox{[\,$\ell {\negdbltinyspace}\in{\negdbltinyspace} \Dbl{N}$\,]}
\,.
\end{equation}
Applying the inverse star operation (\ref{Inv_StarOpPsiModeExpansion}) to the Hamiltonian 
\rf{PureDT_HamiltonianModeExpansion} leads to 
\begin{equation}\label{pureDT_HamiltonianModeExpansionStarW}
\Hop
\,=\,
-\,\sqrt{2{\tinyspace}\GG} \,{\dbltinyspace}
 \overline{W}^{{\dbltinyspace}(3)*}_{{\negqdrpltinyspace}-2}
\,+\, Y^*
\,,
\end{equation}
where 
$\overline{W}^{{\dbltinyspace}(3)}_{{\negqdrpltinyspace}n}$
[{\tinyspace}$n \!\in\! \Dbl{Z}${\trpltinyspace}]
denotes the two-reduced $W^{(3)}$ operators, which are defined by 
\begin{equation}\label{2reducedW3operator}
\overline{W}^{{\dbltinyspace}(3)}_{{\negqdrpltinyspace}n}
\,\define\,
\quarter \biggl( \, \onethird{\negtrpltinyspace}
\sum_{k+l+m
      {\tinyspace}={\tinyspace}2n}
{\negtrpltinyspace}{\negtrpltinyspace}{\negtrpltinyspace}
 : \a_k\a_l\a_m : 
\,+\,{\trpltinyspace}
\quarter{\tinyspace} \a_{2n} \, \biggr)
\,,
\qquad
\mbox{[\,$n$, $k$, $l$, $m {\negdbltinyspace}\in{\negdbltinyspace} \Dbl{Z}$\,]}
\,,
\end{equation}
and $Y$ is introduced so as to satisfy the no-big-bang condition \rf{NoBigBangCondition}, and it has the form
\begin{equation}\label{PureDT_Y}
Y
\,=\,
\sqrt{\frac{2}{\GG}}
\Big({\dbltinyspace}
  \frac{\a_6}{2}{\tinyspace}
  {\tinyspace}-{\tinyspace}
  \mu_2{\tinyspace}
  \a_2
{\negtinyspace}\Big)
\,.
\end{equation}
The $Y$ term arises from the nonzero expectation values of 
$\sqrt{\GG/2}{\trpltinyspace} \phi_n$, i.e.,\ 
$\alpha_{-n} / n$
[\,$n {\negdbltinyspace}\in{\negdbltinyspace} \Dbl{N}$\,].

Equations \eqref{pureDT_HamiltonianModeExpansionStarW}--\eqref{PureDT_Y}  specify the precise sense in which the two-reduced $W^{(3)}$ representation enters the present formulation: the Hamiltonian is expressed in terms of the particular mode $W_{-2}^{(3)}$ in this representation.
In the derivation of the SD equations below, we use the explicit Hamiltonian together with the oscillator commutation relations, rather than the commutation relations of the $W^{(3)}$  algebra itself.

\newcommand{\TwoReducedCond}{%
(The following point will be omitted.)
It is known that the following conditions are satisfied, 
\begin{equation}\label{PureDT_VanishingEvenMode}
\pder{j_{2n}} Z[j]  \,=\,  0
\,,
\qquad
\mbox{[\,$n {\negdbltinyspace}\in{\negdbltinyspace} \Dbl{N}$\,]}
\,.
\end{equation}
}%

\subsubsection{Amplitudes and Schwinger-Dyson equation}
\label{sec:PureDTamplitudes}

The amplitudes are obtained from derivatives of the generating function $Z[j]$ in \rf{GeneratingFunModeExpansion} 
with respect to $j_\ell$ 
and are defined by 
\begin{equation}\label{GeneralAmplitudes}
f_\NN(\ell_1,\ldots,\ell_\NN)
\ \define \ 
\lim_{\T\rightarrow\infty}
  \vac {\tinyspace} \E^{-T \Hop}%
    \phi^\dagger_{\ell_1} \dots \phi^\dagger_{\ell_\NN}
  \cuum
\,.
\end{equation}
To make the notion of connected amplitudes explicit, let us also introduce the connected generating functional
$
W[j]:=\log Z[j].
$
Then the connected amplitudes are equivalently generated by derivatives of $W[j]$, and
we denote them by
\begin{equation}\label{GeneralConnectedAmplitudes}
f_\NN^{{\rm conn}}(\ell_1,\ldots,\ell_\NN)
\ \define \ 
\lim_{\T\rightarrow\infty}
  \vac {\tinyspace} \E^{-T \Hop}%
    \phi^\dagger_{\ell_1} \dots \phi^\dagger_{\ell_\NN}
  \cuum^{{\rm conn}}
\,,
\end{equation}
which are related to \rf{GeneralAmplitudes} by the cumulant expansions
\begin{eqnarray}
f_1(\ell_1)
\!\!&=&\!\!
f_1^{{\rm conn}}(\ell_1)
\,,
\\
f_2(\ell_1,\ell_2)
\!\!&=&\!\!
f_2^{{\rm conn}}(\ell_1,\ell_2)
+
f_1^{{\rm conn}}(\ell_1)
f_1^{{\rm conn}}(\ell_2)
\,,
\\
f_3(\ell_1,\ell_2,\ell_3)
\!\!&=&\!\!
f_3^{{\rm conn}}(\ell_1,\ell_2,\ell_3)
+
f_1^{{\rm conn}}(\ell_1)
f_2^{{\rm conn}}(\ell_2,\ell_3)
\nonumber
\\
&&\!\!
+\>
f_1^{{\rm conn}}(\ell_2)
f_2^{{\rm conn}}(\ell_3,\ell_1)
+
f_1^{{\rm conn}}(\ell_3)
f_2^{{\rm conn}}(\ell_1,\ell_2)
\nonumber
\\
&&\!\!
+\>
f_1^{{\rm conn}}(\ell_1)
f_1^{{\rm conn}}(\ell_2)
f_1^{{\rm conn}}(\ell_3)
\,.
\end{eqnarray}
In general, we have 
\begin{equation}\label{RelationGeneralConnectedAmplitudes}
f_\NN(\ell_1,\ldots,\ell_\NN)
\ = \ 
\sum_{%
\mbox{\scriptsize all possible $\{ A^{(i)} \}$
}
}
 \prod_i
  f_{\NN_i}^{{\rm conn}}(\ell_{\sigma^{(i)}_1},\ldots,\ell_{\sigma^{(i)}_{\NN_i}})
\,,
\end{equation}
where 
$\{{\negtinyspace} A^{(i)} \}$ is a collection of subsets 
$
A^{(i)} = \{ \sigma^{(i)}_1,\, \ldots, \, \sigma^{(i)}_{\NN_i} \}
$  
of $\{1,\ldots,N\}$, satisfying
\begin{equation}
{}^\forall{\neghalftinyspace} i, \ 
A^{(i)} \neq \varnothing
\,,\qquad
{}^\forall{\neghalftinyspace} i \neq {}^\forall{\negdbltinyspace} j, \ 
A^{(i)} {\negtinyspace}\cap{\negtinyspace} A^{(j)} = \varnothing
\,,\qquad
\bigcup_i A^{(i)} = \{1,\,\ldots,\,\NN\}
\,.
\end{equation}

With the formal large-$T$ stationarity prescription specified after \eqref{GeneratingFunModeExpansion}, the quantity
\begin{equation}\label{GeneratingFunModeExpansionSDequation0}
  \pder{{\tinyspace}T}
  \vac {\tinyspace} \E^{-T \Hop} 
    \exp{\negtrpltinyspace}\bigg(
      \sum_{\ell=1}^\infty \phi^\dagger_\ell j_\ell
    {\negtinyspace}\bigg)
  {\negtinyspace}\cuum
\,,
\end{equation}
is required to vanish in the $T\to\infty$ limit. Using the no-big-bang condition \eqref{NoBigBangCondition}, we then obtain an SD equation of the form
\begin{equation}\label{GeneratingFunModeExpansionSDequation}
0  \ = \ 
\lim_{\T\rightarrow\infty}
  \vac {\tinyspace} \E^{-T \Hop} 
   \commutatorBig{\Hop}{
    \exp{\negtrpltinyspace}\bigg(
      \sum_{\ell=1}^\infty \phi^\dagger_\ell j_\ell
    {\negtinyspace}\bigg)
   \!}
  {\negtinyspace}\cuum
\,.
\end{equation}
Differentiating equation \rf{GeneratingFunModeExpansionSDequation} with respect to $j_{\ell_1}, \ldots,  j_{\ell_N}$,
and then setting all sources to zero, we find 
\begin{align}
0&=
\frac{\partial^N}{\partial j_{\ell_1} \ldots \partial j_{\ell_N}}
\lim_{T \to \infty}
  \vac {\tinyspace} \E^{-T \Hop}%
   \commutatorBig{\Hop}{
    \exp{\negtrpltinyspace}\bigg(
      \sum_{\ell=1}^\infty \phi^\dagger_\ell j_\ell
    {\negtinyspace}\bigg)
   \!}
  {\negtinyspace}\cuum
\Big|_{j_1 = j_2 = j_3 = \ldots = 0}
\nonumber\\
&=
\lim_{T \to \infty}
  \vac {\tinyspace} \E^{-T \Hop}%
   \commutator{\Hop}{\phi^\dagger_{\ell_1} \cdots \phi^\dagger_{\ell_N}}
  \cuum
\nonumber\\
&=
\sum_{i=1}^N
\lim_{T \to \infty}
\vac {\tinyspace} \E^{-T \Hop}%
\phi^\dagger_{\ell_1} \cdots \phi^\dagger_{\ell_{i-1}}
\commutator{\Hop}{\phi^\dagger_{\ell_i}}
\phi^\dagger_{\ell_{i+1}} \cdots \phi^\dagger_{\ell_{N}}
\cuum
\,.
\label{sd_eq}
\end{align}

Here we introduce the Laplace transforms of the amplitudes \eqref{GeneralAmplitudes} and 
of the connected amplitudes \eqref{GeneralConnectedAmplitudes}, defined by
\begin{align}
\tilde{f}_\NN(\xi_1,\ldots,\xi_\NN)&=
\sum_{\ell_1=1}^\infty
\ldots
\sum_{\ell_\NN=1}^\infty
\xi_1^{- \ell_1/2 - 1}
\ldots
\xi_\NN^{- \ell_\NN/2 - 1}
f_\NN(\ell_1,\ldots,\ell_\NN)
\,,
\label{laplace_amp}
\\
\tilde{f}_\NN^{{\rm conn}}(\xi_1,\ldots,\xi_\NN)&=
\sum_{\ell_1=1}^\infty
\ldots
\sum_{\ell_\NN=1}^\infty
\xi_1^{- \ell_1/2 - 1}
\ldots
\xi_\NN^{- \ell_\NN/2 - 1}
f_\NN^{{\rm conn}}(\ell_1,\ldots,\ell_\NN)
\,.
\label{laplace_amp_conn}
\end{align}
We also introduce
\begin{align}
\widetilde{\calF}_\NN^{{\rm conn}}(\xi_1,\ldots,\xi_\NN)&=
\Omega_1(\xi_1)\, \delta_{N,1}
+
\GG\, \Omega_2(\xi_1,\xi_2)\, \delta_{N,2}
+
\tilde{f}_\NN^{{\rm conn}}(\xi_1,\ldots,\xi_\NN)
\,,
\label{fr_energy}
\end{align}
where, for pure DT, 
\begin{align}
\Omega_1(\xi_1)
&=
\xi_1^{3/2} - \cc_2\, \xi_1^{-1/2}
\,,
\label{omega1}
\\
\Omega_2(\xi_1,\xi_2)
&=
\frac{1}
     {2 \sqrt{\xi_1 \xi_2}{\tinyspace}
      \big({\negtinyspace}
        \sqrt{\xi_1} +{\negtrehalftinyspace} \sqrt{\xi_2}
      {\qdrpltinyspace}\big)^{{\negtinyspace}2}}\,.
\label{omega2}
\end{align}
Then, the SD equation \eqref{sd_eq} for the Hamiltonian \eqref{PureDT_HamiltonianModeExpansion} of pure DT 
is shown to yield (see Section 4.1.2 and 4.1.3 of \cite{FMW2025a} or Section \ref{sec:MultiDTamplitudes} in this paper for details),
\begin{align}
0&=
\sum_{i=1}^{N}
\pder{\xi_i}\bigg(
\widetilde{\calF}_{N+1}^{{\rm conn}}(\xi_i,\bm{\xi}_{I})
+ 
\sum_{I_1 \cup I_2=I \backslash \{i\}}
\widetilde{\calF}_{|I_1|+1}^{{\rm conn}}(\xi_i,\bm{\xi}_{I_1})\,
\widetilde{\calF}_{|I_2|+1}^{{\rm conn}}(\xi_i,\bm{\xi}_{I_2})
\nonumber\\
&\hspace{5em}
- \left(\xi_i^3-2\cc_2 \xi_i\right) \delta_{N,1}
+ 2\, \GG 
\sum_{\substack{j=1 \\ (j \neq i)}}^{\NN}
\pder{\xi_j}
\frac{\widetilde{\calF}_{N-1}^{{\rm conn}}(\bm{\xi}_{I \backslash \{j\}})
-\widetilde{\calF}_{N-1}^{{\rm conn}}(\bm{\xi}_{I \backslash \{i\}})}
{\xi_i - \xi_j}\bigg)
\,,
\label{sd_pure_dt}
\end{align}
where $\bm{\xi}_{I}=\{\xi_1, \ldots, \xi_N\}$ is defined for the index set $I=\{1, \ldots, N\}$, and $I_1$ and $I_2$ are disjoint subsets of $I \backslash \{i\}$. 
It has been shown \cite{FMW2025a} that 
the CEO topological recursion provides perturbative solutions to equation \eqref{sd_pure_dt}, expanded around $\GG=0$.

\subsection{Multicritical DT}
\label{sec:MultiDT}

\subsubsection{Background}\label{sec:MultiDT_bg}

Before turning to the formulation of the $m$-th multicritical DT, let us briefly recall
the theoretical background relevant to this case. Multicritical DT is obtained by
extending pure DT to higher critical points, and in the continuum limit it is related to
Liouville gravity coupled to the $(2,2m-1)$ minimal model.

In the previous literature, the multicritical theory was mainly understood through the
continuum limit of matrix models. In particular, a continuum string-field-theoretic
description reproducing the loop equations of multicritical matrix models was developed
in \cite{Ishibashi:1993nq}, while the corresponding multicritical transfer-matrix
formulation and its relation to Liouville gravity were analyzed in \cite{SFT:GK,Moore:1991ir}.

The continuum Hamiltonian of the $m$-th multicritical DT was then constructed by
applying the string-field-theoretic framework developed in \cite{SFT:Watabiki} to the
multicritical transfer-matrix formulation studied in the matrix-model approach. In this
framework, the multicritical string field theory is formulated directly at the level of the
continuum Hamiltonian, and 
the multicritical Hamiltonian can be represented in terms of the same $W_{-2}^{(3)}$ mode in the two-reduced $W^{(3)}$ representation
after rewriting the theory in terms of mode expansions
\cite{SFT:AW}.

What had not been made equally explicit, however, was the direct derivation of the SD equations from this Hamiltonian, together with the construction of a one-cut disk solution and of the perturbative amplitudes around it by the CEO topological recursion. In this section, we derive these equations explicitly and construct the corresponding spectral curve.
This perspective is also useful for clarifying two further aspects of the multicritical
theory. First, for a distinguished specialization of the cosmological constants, one
obtains the so-called conformal background, in which the disk amplitude coincides with
that of the $(2,2m-1)$ minimal string. Second, after an appropriate specialization and
normalization, the large-$m$ limit of the disk amplitude reproduces the Laplace dual of the JT-gravity disk
amplitude. These points will be discussed in Section~\ref{sec:MultiDTandJTgravity}.

\subsubsection{Formulation}
\label{sec:MultiDTdef}

Multicritical DT is an extension of pure DT
and consists of $n$-gons with 
$n{\negtrpltinyspace}={\negtrpltinyspace}3$, 
$4$, \ldots. 
In this model, the critical point of the DT partition function is chosen to be of order $m$ \cite{MM:Kaza}.
Pure DT, as discussed in Section \ref{sec:PureDT}, 
is equivalent to multicritical DT with 
$m{\negtrpltinyspace}={\negtrpltinyspace}2$. 
According to \cite{SFT:AW}, 
a continuum theory of multicritical DT 
can also be formulated using mode expansions.
The following is a brief summary of their results in the multicritical case.

In the continuum limit,
the generating function for multicritical DT 
is given by \rf{GeneratingFunModeExpansion},
together with the commutation relation \rf{CommutationRelationPhi}
and the vacuum condition \rf{vacuumCondition_phi}.
The only difference from the pure DT case lies in the Hamiltonian.
To obtain pure DT as a special case, 
one performs a shift of the operators
$\alpha_n$ (i.e., $\phi_n$ for $n = -1$ and $-5$)
appearing in the two-reduced $W^{(3)}$ operator. 
The Hamiltonian before this shift is 
\begin{eqnarray}\label{OriginalDT_HamiltonianModeExpansion}
\Hop^{({\rm origin})} 
\!\!&\define&\!\!
-\,\sqrt{2{\tinyspace}\GG} \,{\dbltinyspace}
 \overline{W}^{{\dbltinyspace}(3)*}_{{\negqdrpltinyspace-}2}
\Big|_{\nu=0,\,\,\lambda_n = 0\,[n\in\dblsmall{N}{\dbltinyspace}]}
\nonumber\\&=&\!\!
   -{\qntpltinyspace}
       \frac{\GG}{4}{\tinyspace} \phi_{{\halftinyspace}4}
 \,-\, \frac{\GG^{{\tinyspace}2}}{4}{\dbltinyspace}
       \phi_1^2 {\tinyspace} \phi_{{\tinyspace}2}
 \,-\,
  \frac{1}{2}
  \sum_{\ell=6}^\infty {\tinyspace} \sum_{n=1}^{\ell-5}
  \phi_n^\dagger \phi_{\ell-n-4}^\dagger {\tinyspace}
  \ell \phi_\ell
\nonumber\\&&\!\!
 -\>
  \frac{\GG}{4}
  \sum_{\ell=1}^\infty {\tinyspace} \sum_{n=\max(5-\ell,1)}^\infty\!\!
  \phi_{n+\ell-4}^\dagger {\tinyspace}
  n {\halftinyspace} \phi_n {\tinyspace} \ell \phi_\ell
\,.
\end{eqnarray}
The general shift of the two-reduced $W^{(3)}$ operator is implemented by 
\begin{eqnarray}\label{CoherentEigenValuesMultiDT}
&&\hspace{-16pt}%
\nu \define 0
\,,
\quad
\lambda_{{\tinyspace}2i+1} \define 
-\,
\frac{(-1)^{m-i} \cc_{m-i}}
     {2{\tinyspace}i{\negdbltinyspace}+{\negdbltinyspace}1}
      \sqrt{\frac{2}{\GG}}
\ \ \mbox{[{\dbltinyspace}%
$i{\negdbltinyspace}={\negdbltinyspace}0$, \ldots, 
$m{\negtrpltinyspace}-{\negtrpltinyspace}2$%
{\dbltinyspace}]}
\,,
\quad
\lambda_{{\tinyspace}2m+1} \define 
\frac{1}
     {2{\tinyspace}m{\negdbltinyspace}+{\negdbltinyspace}1}
      \sqrt{\frac{2}{\GG}}
\,,
\nonumber\\
&&\hspace{-16pt}%
\lambda_{{\tinyspace}n} \define 0
\ \ 
\mbox{[\,$n {\negdbltinyspace}\in{\negdbltinyspace} \Dbl{N}$, 
         $n {\negtrpltinyspace}\neq{\negtrpltinyspace} 1$, $3$, 
           \ldots,
           $2{\tinyspace}m{\negtrpltinyspace}-{\negtrpltinyspace}5$,
           $2{\tinyspace}m{\negtrpltinyspace}-{\negtrpltinyspace}3$,
           $2{\tinyspace}m{\negtrpltinyspace}+{\negtrpltinyspace}1$
     \!]}
\,.
\end{eqnarray}
The Hamiltonian is given by 
\rf{pureDT_HamiltonianModeExpansionStarW} 
under the shift
\rf{DefAlphaOperatorPureDT} with $p=2$ together with \rf{CoherentEigenValuesMultiDT},
where the operator $Y^*$ is a polynomial in the creation operators 
$\phi_n^\dagger$, chosen such that the Hamiltonian 
$\Hop$ satisfies the no-big-bang condition \rf{NoBigBangCondition}.
We obtain 
the $m$-th multicritical DT defined by the Hamiltonian $\Hop$:
\begin{eqnarray}\label{MultiDT_HamiltonianModeExpansion}
\Hop \!\!&=&\!\!
\Hop^{({\rm origin})}%
\Big|_{\phi_\ell {\tinyspace}\to{\tinyspace}
       \phi_\ell {\tinyspace}+{\tinyspace}
       \lambda_\ell \sqrt{2/\GG}}
{\dbltinyspace}+
Y^*
\nonumber\\&=&\!\!
   -{\qntpltinyspace}
       \frac{\GG}{4}{\tinyspace} \phi_{{\halftinyspace}4}
 \,-\, \Big(
         (-1)^m \cc_m
         -
         \frac{\GG}{2}{\dbltinyspace}\phi_1
       \Big)^{{\negtrpltinyspace}2}
       \phi_{{\tinyspace}2}
\nonumber\\&&\!\!
 -\>
   \sum_{\ell=1}^\infty
     \phi_{\ell+2m-3}^\dagger {\dbltinyspace} \ell \phi_\ell
 \,+\,
   \sum_{i=2}^{m} {\tinyspace} \sum_{\ell=\max(2i-2m+4,1)}^\infty\!\!
     (-1)^i \cc_i {\tinyspace} \phi_{\ell+2m-2i-3}^\dagger
     {\dbltinyspace} \ell \phi_\ell
\nonumber\\&&\!\!
 -\> \frac{1}{2}
  \sum_{\ell=6}^\infty {\tinyspace} \sum_{n=1}^{\ell-5}
    \phi_n^\dagger \phi_{\ell-n-4}^\dagger
    {\dbltinyspace} \ell \phi_\ell
\,-\,
  \frac{\GG}{4}
  \sum_{\ell=1}^\infty {\tinyspace} \sum_{n=\max(5-\ell,1)}^\infty\!\!
    \phi_{n+\ell-4}^\dagger
    {\dbltinyspace} n {\halftinyspace} \phi_n
    {\dbltinyspace} \ell \phi_\ell
\,,
\end{eqnarray}
where 
$\cc_i$ 
[{\tinyspace}$i{\negtrpltinyspace}={\negtrpltinyspace}2$,
 \ldots, $m${\tinyspace}]
are the cosmological constants with mass dimension $i$, 
and $\GG$ counts the number of handles of a 2D surface as before.

\subsubsection{Amplitudes and Schwinger-Dyson equation}
\label{sec:MultiDTamplitudes}

For the Hamiltonian $H$ in \eqref{MultiDT_HamiltonianModeExpansion}, the amplitudes and the connected amplitudes are defined by \eqref{GeneralAmplitudes} and \eqref{GeneralConnectedAmplitudes}, respectively. We derive the corresponding SD equations directly from \eqref{sd_eq}. In order to compute these amplitudes, we first evaluate the commutator for the unshifted Hamiltonian $H^{(\mathrm{origin})}$ in \eqref{OriginalDT_HamiltonianModeExpansion}.

For the unshifted Hamiltonian $\Hop^{({\rm origin})}$ in \eqref{OriginalDT_HamiltonianModeExpansion}, we have
\begin{align}
\commutator{\Hop^{({\rm origin})}}{\phi^\dagger_{\ell_i}}&=
-\frac{\GG^2}{2} \delta_{\ell,1}\, \phi_1 \phi_2
-\frac{\GG^2}{4} \delta_{\ell,2}\, \phi_1^2 
- \frac{\GG}{4} \delta_{\ell, 4}
\nonumber
\\
&\ \ \
-\frac12 \ell\, \sum_{n=1}^{\ell-5} \phi_n^{\dagger} \phi_{\ell-n-4}^{\dagger}
-\frac{\GG}{2} \ell\, \sum_{n = \mathrm{max}(5-\ell,1)}^{\infty} 
n\, \phi_{n+\ell-4}^{\dagger} \phi_n
\,,
\label{com_H_phi}
\end{align}
and hence we obtain
\begin{align}
&
-\phi_{\ell_1}^{\dagger}\cdots \phi_{\ell_{i-1}}^{\dagger}
\commutator{\Hop^{({\rm origin})}}{\phi^\dagger_{\ell_i}}
\phi_{\ell_{i+1}}^{\dagger}\cdots \phi_{\ell_N}^{\dagger}
\nonumber\\
&=
\phi_{\ell_1}^{\dagger}\cdots \breve{\phi}_{\ell_i}^{\dagger}\cdots
\phi_{\ell_N}^{\dagger}
\left(\frac{\GG^2}{2}\delta_{\ell_i,1}\, \phi_1 \phi_2
+\frac{\GG^2}{4} \delta_{\ell_i,2}\, \phi_1^2 
+ \frac{\GG}{4} \delta_{\ell_i, 4}
\right)
\nonumber\\
&\ \
+\phi_{\ell_1}^{\dagger}\cdots \breve{\phi}_{\ell_i}^{\dagger}\cdots
\phi_{\ell_N}^{\dagger}\left(
\frac12 \ell_i \sum_{n=1}^{\ell_i-5} \phi_n^{\dagger} \phi_{\ell_i-n-4}^{\dagger}
+ \frac{\GG}{2} \ell_i 
\sum_{n = \mathrm{max}(5-\ell_i,1)}^{\infty} n\, \phi_{n+\ell_i-4}^{\dagger} \phi_n
\right)
\nonumber\\
&\ \
+\frac{\GG^2}{2} \sum_{j=i+1}^N
\phi_{\ell_1}^{\dagger}\cdots \breve{\phi}_{\ell_i}^{\dagger} \cdots
\breve{\phi}_{\ell_j}^{\dagger}\cdots
\phi_{\ell_N}^{\dagger}
\left(
\delta_{\ell_i+\ell_j,2}\, \phi_2 + \delta_{\ell_i+\ell_j,3}\, \phi_1
\right)
\nonumber\\
&\ \
+\frac{\GG^2}{2} \sum_{i+1 \le j<k \le N}
\delta_{\ell_i+\ell_j+\ell_k, 4}\,
\phi_{\ell_1}^{\dagger}\cdots \breve{\phi}_{\ell_i}^{\dagger}
\cdots \breve{\phi}_{\ell_j}^{\dagger}\cdots
\breve{\phi}_{\ell_k}^{\dagger}\cdots \phi_{\ell_N}^{\dagger}
\nonumber\\
&\ \
+\frac{\GG}{2} \sum_{j=i+1}^N
\ell_i\, \ell_j\, \theta_{\ell_i+\ell_j-4,1}\,
\phi_{\ell_i+\ell_j-4}^{\dagger}\,
\phi_{\ell_1}^{\dagger}\cdots \breve{\phi}_{\ell_i}^{\dagger}
\cdots \breve{\phi}_{\ell_j}^{\dagger} \cdots \phi_{\ell_N}^{\dagger}
\,.
\label{com_H_phi_multi}
\end{align}
Here $\breve{\phi}_{\ell}^{\dagger}$ denotes that  $\phi_{\ell}^{\dagger}$ is omitted, 
and $\theta_{k,\ell}=1$ if $k \ge \ell$ and $\theta_{k,\ell}=0$ if $k< \ell$. 
Using the commutation relation \eqref{com_H_phi_multi} together with 
the SD equation \eqref{sd_eq}, 
we find that the Laplace-transformed amplitudes \eqref{laplace_amp} for the Hamiltonian $H$ with the shift specified in \eqref{MultiDT_HamiltonianModeExpansion} satisfy
\begin{align}
0&=
\sum_{i=1}^N \Biggl[\pder{\xi_i}
\tilde{f}_{\NN+1}(\xi_i,\bm{\xi}_{I})
+2 \pder{\xi_i}
\left(
\xi_i^{m-1/2} 
\tilde{f}_{\NN}(\bm{\xi}_I)-
\left(\xi_i^{m-1/2} 
\tilde{f}_{\NN}(\bm{\xi}_I)\right)_{\mathrm{reg}(\xi_i)}
\right)
\nonumber\\
&\hspace{3em}
-2 \sum_{p=0}^{m-2} (-1)^{m-p} \mu_{m-p} \pder{\xi_i}
\left(
\xi_i^{p-1/2} 
\tilde{f}_{\NN}(\bm{\xi}_I)-
\left(\xi_i^{p-1/2} 
\tilde{f}_{\NN}(\bm{\xi}_I)\right)_{\mathrm{reg}(\xi_i)}
\right)
\nonumber\\
&\hspace{3em}
+\GG \sum_{\substack{j=1 \\ (j \neq i)}}^{\NN}
\frac{\partial^2}{\partial \xi_i \partial \xi_j}
\frac{\left(1+\xi_i^{-1/2}\xi_j^{1/2}\right) 
\tilde{f}_{\NN-1}(\bm{\xi}_{I \backslash \{j\}})
-\left(\xi_i^{1/2}\xi_j^{-1/2}+1\right) 
\tilde{f}_{\NN-1}(\bm{\xi}_{I \backslash \{i\}})}
{\xi_i-\xi_j}
\nonumber\\
&\hspace{3em}
-\frac14 \left(\GG  \xi_i^{-3}
+ 4 \mu_m^2 \xi_i^{-2}\right)
\tilde{f}_{\NN-1}(\bm{\xi}_{I \backslash \{i\}})
+ (-1)^m \mu_m \GG \sum_{\substack{j=1 \\ (j \neq i)}}^{\NN}
\xi_i^{-2}\xi_j^{-3/2}\,
\tilde{f}_{\NN-2}(\bm{\xi}_{I \backslash \{i, j\}})
\nonumber\\
&\hspace{3em}
-\frac12 \GG^2 \sum_{\substack{1 \le j<k \le N \\ (j,k \ne i)}}
\xi_i^{-2}\xi_j^{-3/2}\xi_k^{-3/2}\,
\tilde{f}_{\NN-3}(\bm{\xi}_{I \backslash \{i, j, k\}})\Biggr]
\,,
\label{sd_multicritical}
\end{align}
where $\bm{\xi}_I=\{\xi_1, \ldots, \xi_N\}$ is defined for $I=\{1, \ldots, N\}$, 
and $f(\xi)_{\mathrm{reg}(\xi)}$ denotes 
the regular part of a Laurent series $f(\xi)$ of $\xi$.

When we focus on the variable $\xi_1$ in the SD equation \eqref{sd_multicritical}, 
we obtain 
from the Laplace-transformed amplitude \eqref{laplace_amp} that, as $\xi_1 \to \infty$,
\begin{align}
&
\tilde{f}_{\NN+1}(\xi_1,\bm{\xi}_{I})=O(\xi_1^{-3}),
\ \ \ \
\xi_1^{p-1/2} \tilde{f}_{\NN}(\bm{\xi}_I)=O(\xi_1^{p-2}),
\nonumber\\
&
\tilde{f}_{\NN-1}(\bm{\xi}_{I \backslash \{j\}})=O(\xi_1^{-3/2})
\ \ \textrm{for}\ \ j\ne 1,
\nonumber\\
&
\tilde{f}_{\NN-1}(\bm{\xi}_{I \backslash \{1\}})=
\tilde{f}_{\NN-2}(\bm{\xi}_{I \backslash \{1, j\}})=
\tilde{f}_{\NN-3}(\bm{\xi}_{I \backslash \{1, j, k\}})=O(\xi_1^0)
\,.
\end{align}
By taking these asymptotic behaviors into account and 
integrating with respect to $\xi_1$, 
we deduce a separated SD equation for the $m$-th multicritical DT:
\begin{align}
0&=
\tilde{f}_{\NN+1}(\xi_1, \bm{\xi}_{I})
+ 2 \left(
\Omega_1(\xi_1)\, \tilde{f}_{\NN}(\bm{\xi}_{I})-
\left(\Omega_1(\xi_1)\, \tilde{f}_{\NN}(\bm{\xi}_{I})\right)_{\mathrm{reg}(\xi_1)}
\right)
\nonumber\\
&\ \
+2\GG \sum_{i=2}^N \pder{\xi_i}
\frac{\xi_1^{-1/2}\xi_i^{1/2}\, \tilde{f}_{\NN-1}(\bm{\xi}_{I \backslash \{i\}})
- \tilde{f}_{\NN-1}(\bm{\xi}_{I \backslash \{1\}})}
{\xi_1-\xi_i}
\nonumber\\
&\ \
+\frac18 \left(\GG  \xi_1^{-2}+ 8 \mu_m^2 \xi_1^{-1}\right)
\tilde{f}_{\NN-1}(\bm{\xi}_{I \backslash \{1\}})
-(-1)^m \mu_m \GG \sum_{i=2}^N 
\xi_1^{-1}\xi_i^{-3/2}\,
\tilde{f}_{\NN-2}(\bm{\xi}_{I \backslash \{1, i\}})
\nonumber\\
&\ \
+\frac12 \GG^2 \sum_{2 \le i<j \le N}
\xi_1^{-1}\xi_i^{-3/2}\xi_j^{-3/2}\,
\tilde{f}_{\NN-3}(\bm{\xi}_{I \backslash \{1, i, j\}})
+C_{N-1}(\bm{\xi}_{I \backslash \{1\}})
\,,
\label{sp_sd_m_crit_dt}
\end{align}
where $C_{N-1}(\bm{\xi}_{I \backslash \{1\}})$ is a function of $\bm{\xi}_{I \backslash \{1\}}$, 
and we define $\Omega_1(\xi)$ by
\begin{align}
\Omega_1(\xi)=
\xi^{m-1/2}
-\sum_{p=0}^{m-2} (-1)^{m-p} \mu_{m-p}\, \xi^{p-1/2}
\,.
\label{omega1_multi}
\end{align}

We now define amplitudes $\widetilde{\calF}_\NN^{{\rm conn}}(\xi_1,\ldots,\xi_\NN)$ by \eqref{fr_energy}, 
where \eqref{omega1_multi} for $\Omega_1(\xi)$ 
is used as a generalization of \eqref{omega1}.
We note that this function reduces to \eqref{omega1} when $m=2$.
Then, starting from the separated SD equation \eqref{sp_sd_m_crit_dt}, 
we prove the following proposition in Appendix \ref{app:proof_prop}.
(Note that when $m=2$, equation \eqref{sp_sd_m_crit_dt_rev} reduces to 
equation \eqref{sd_pure_dt}.)

\begin{prop}\label{prop:sd_multi_dt}
The amplitudes $\widetilde{\calF}_\NN^{{\rm conn}}(\xi_1,\ldots,\xi_\NN)$, 
determined by equation \eqref{sp_sd_m_crit_dt} and also providing solutions to the SD equation \eqref{sd_multicritical}, 
satisfy
\begin{align}
0&=
\widetilde{\calF}_{N+1}^{{\rm conn}}(\xi_1,\bm{\xi}_{I})
+ 
\sum_{I_1 \cup I_2=I \backslash \{1\}}
\widetilde{\calF}_{|I_1|+1}^{{\rm conn}}(\xi_1,\bm{\xi}_{I_1})\,
\widetilde{\calF}_{|I_2|+1}^{{\rm conn}}(\xi_1,\bm{\xi}_{I_2})
\nonumber\\
&\ \ \ \
+ 2\, \GG {\negdbltinyspace}
\sum_{i=2}^{\NN}
\pder{\xi_i}
\frac{\widetilde{\calF}_{N-1}^{{\rm conn}}(\bm{\xi}_{I \backslash \{i\}})
-\widetilde{\calF}_{N-1}^{{\rm conn}}(\bm{\xi}_{I \backslash \{1\}})}
{\xi_1 - \xi_i}
-2 \left(\Omega_1(\xi_1)\, 
\tilde{f}_{N}^{{\rm conn}}(\bm{\xi}_I)\right)_{\mathrm{reg}(\xi_1)}
\nonumber\\
&\ \ \ \
-\left(\Omega_1(\xi_1)^2\right)_{\mathrm{reg}(\xi_1)}\, \delta_{N,1}
-2\, \GG \pder{\xi_2} 
\frac{\left(\xi_1^{-1/2}\xi_2^{1/2}\Omega_1(\xi_1)\right)_{\mathrm{reg}(\xi_1)}-\Omega_1(\xi_2)_{\mathrm{reg}(\xi_2)}}
{\xi_1-\xi_2}\, \delta_{N,2}
\nonumber\\
&\ \ \ \
+\tilde{C}_{N-1}(\bm{\xi}_{I \backslash \{1\}})
\,,
\label{sp_sd_m_crit_dt_rev}
\end{align}
where $\tilde{C}_{N-1}(\bm{\xi}_{I \backslash \{1\}})$ is a function of $\bm{\xi}_{I \backslash \{1\}}$.
\end{prop}

\subsubsection{Perturbative amplitudes and topological recursion}
\label{sec:TR_DT}

In the following, we consider formal perturbative solutions of equation \eqref{sp_sd_m_crit_dt_rev} expanded around $\GG=0$, 
\begin{align}
\tilde{f}_\NN^{{\rm conn}}(\xi_1,\ldots,\xi_\NN)&=
\sum_{h=0}^{\infty} \GG^{h+N-1} \tilde{f}_{N}^{{\rm conn}(h)}(\xi_1,\ldots,\xi_\NN)\,,
\label{pert_exp_f}
\\
\widetilde{\calF}_\NN^{{\rm conn}}(\xi_1,\ldots,\xi_\NN)&=
\sum_{h=0}^{\infty} \GG^{h+N-1} \widetilde{\calF}_{N}^{{\rm conn}(h)}(\xi_1,\ldots,\xi_\NN)\,.
\label{pert_exp}
\end{align}
Here, 
since $\GG$ counts the number of handles, this expansion provides the genus expansion.
In particular,  $\tilde{f}_{1}^{{\rm conn}(0)}(\xi_1)$ (or $\widetilde{\calF}_{1}^{{\rm conn}(0)}(\xi_1)$) and 
$\tilde{f}_{2}^{{\rm conn}(0)}(\xi_1, \xi_2)$ (or $\widetilde{\calF}_{2}^{{\rm conn}(0)}(\xi_1, \xi_2)$) give the disk and the cylinder amplitudes, respectively.  

\subsubsection*{Disk amplitude}

For $N=1$, equation \eqref{sp_sd_m_crit_dt_rev} reads
\begin{align}
0=
\widetilde{\calF}_{2}^{\rm conn}(\xi,\xi)
+ 
\widetilde{\calF}_{1}^{\rm conn}(\xi)^2
-2 \left(\Omega_1(\xi)\, 
\tilde{f}_{1}^{{\rm conn}}(\xi)\right)_{\mathrm{reg}(\xi)}
-\left(\Omega_1(\xi)^2\right)_{\mathrm{reg}(\xi)}
+\tilde{C}_{0}
\,.
\label{sd_m_crit_dt_1}
\end{align}
Using the perturbative expansions \eqref{pert_exp_f} and \eqref{pert_exp} of amplitudes, the leading part of equation \eqref{sd_m_crit_dt_1} yields
\begin{align}
\widetilde{\calF}_{1}^{{\rm conn}(0)}(\xi)^2=
\left(\Omega_1(\xi)^2\right)_{\mathrm{reg}(\xi)}
+2 \left(\Omega_1(\xi)\, 
\tilde{f}_{1}^{{\rm conn}(0)}(\xi)\right)_{\mathrm{reg}(\xi)}
+\tilde{C}_{0}^{(0)}
\,,
\label{disk_eq_mdt}
\end{align}
where $\tilde{C}_{0}^{(0)}$ depends on the cosmological constants
$\mu_p$ but not on $\xi$.

Equation \eqref{disk_eq_mdt} is the disk SD equation derived from the multicritical DT Hamiltonian. Although we do not claim that this equation uniquely excludes other possible analytic structures, it admits a one-cut solution. We focus on this solution because the disk amplitude in the continuum limit of multicritical one-matrix models is described by the corresponding one-cut branch \cite{MM:Kaza}. 
It takes the form
\begin{align}
\widetilde{\calF}_{1}^{{\rm conn}(0)}(\xi)&=
\left(\xi^{m-1}+\sum_{p=1}^{m-1}t_{m-p}(\bm{\mu})\, \xi^{p-1} \right)
\sqrt{\xi + \alpha(\bm{\mu})}
=:
M(\xi)\, \sqrt{\sigma(\xi)}\,.
\label{disk_amp_mdt}
\end{align}
Substituting this one-cut solution into equation \eqref{disk_eq_mdt}, together with the required large-$\xi$ behavior, determines the functions introduced below.
In particular, the one-cut data are fixed by the Hamiltonian couplings $\bm{\mu}$ through the equations below and are not introduced as independent parameters.
For later convenience, we write
\begin{align}
M(\xi)=\xi^{m-1}+\sum_{p=1}^{m-1}t_{m-p}(\bm{\mu})\, \xi^{p-1}\,,
\quad
\sigma(\xi)=\xi + \alpha(\bm{\mu})\,.
\end{align}
Here, the $m$ functions $\alpha=\alpha(\bm{\mu})$, $t_1=t_1(\bm{\mu}), \ldots, t_{m-1}=t_{m-1}(\bm{\mu})$, depending on 
$\bm{\mu}=\{\mu_2, \ldots, \mu_{m}\}$ are determined by the following $m$ equations:
\begin{align}
\sum_{\substack{a,b=0 \\ (a+b=p)}}^{m-1} t_a t_b 
+ \alpha \sum_{\substack{a,b=0 \\ (a+b=p-1)}}^{m-1} t_a t_b
=
(-1)^p\left(
\sum_{\substack{a,b=2 \\ (a+b=p)}}^{m} \mu_a \mu_b - 2\mu_p \right),
\quad
p=1, \ldots, m\,,
\label{sp_det_eq}
\end{align}
where $t_0 \equiv 1$ and $\mu_1 \equiv 0$.

For example, when $m=2$, corresponding to the pure DT, equations \eqref{sp_det_eq} reduce to
$$
2t_1 + \alpha=0\,,\quad
t_1^2 + 2\alpha t_1 = -2\mu_2=-\frac34 \mu\,,
$$
and have solutions $t_1=-\sqrt{\mu}/2$ and $\alpha=\sqrt{\mu}$. 
The disk amplitude \eqref{disk_amp_mdt} is then given by
\begin{align}
\widetilde{\calF}_{1}^{{\rm conn}(0)}(\xi)=\left(\xi - \frac{\sqrt{\mu}}{2}\right)\sqrt{\xi+\sqrt{\mu}}\,.
\label{m2_dt_sp}
\end{align}

When $m=3$, equations \eqref{sp_det_eq} become
$$
2t_1 + \alpha=0\,,\quad
2t_2 + t_1^2 + 2\alpha t_1 = -2\mu_2\,,\quad
2t_1 t_2 + \alpha (2t_2 + t_1^2) = 2\mu_3\,.
$$
In particular, by taking $\mu_2 = 5\mu/8$ and $\mu_3 = 0$, 
which corresponds to the ``conformal background'' \cite{Moore:1991ir,SFT:GK} 
(see Section \ref{sec:MultiDTandJTgravity}), 
these equations admit the solutions 
$t_1=-\sqrt{\mu}/2$, $t_2=-\mu/4$, and $\alpha=\sqrt{\mu}$. 
In this case, the disk amplitude \eqref{disk_amp_mdt} is \cite{SFT:GK},
\begin{align}
\widetilde{\calF}_{1}^{{\rm conn}(0)}(\xi)=\left(\xi^2 - \frac{\sqrt{\mu}}{2}\, \xi - \frac{\mu}{4}
\right)\sqrt{\xi+\sqrt{\mu}}\,.
\label{m3_dt_sp}
\end{align}

\subsubsection*{Cylinder amplitude}

For $N=2$, equation \eqref{sp_sd_m_crit_dt_rev} reads
\begin{align}
0&=
\widetilde{\calF}_{3}^{\rm conn}(\xi_1,\xi_1, \xi_2)
+ 
2 \widetilde{\calF}_{1}^{\rm conn}(\xi_1)\, \widetilde{\calF}_{2}^{\rm conn}(\xi_1,\xi_2)
\nonumber\\
&\ \ \ \
+ 2\, \GG {\negdbltinyspace}
\pder{\xi_2}
\frac{\widetilde{\calF}_{1}^{\rm conn}(\xi_1) - \widetilde{\calF}_{1}^{\rm conn}(\xi_2)}{\xi_1 - \xi_2}
-2 \left(\Omega_1(\xi_1)\, 
\tilde{f}_{2}^{{\rm conn}}(\xi_1, \xi_2)\right)_{\mathrm{reg}(\xi_1)}
\nonumber\\
&\ \ \ \
-2\, \GG \pder{\xi_2} 
\frac{\left(\xi_1^{-1/2}\xi_2^{1/2}\Omega_1(\xi_1)\right)_{\mathrm{reg}(\xi_1)}-\Omega_1(\xi_2)_{\mathrm{reg}(\xi_2)}}
{\xi_1-\xi_2}
+\tilde{C}_{1}(\xi_2)
\,.
\label{sd_m_crit_dt_2}
\end{align}
Using the perturbative expansions \eqref{pert_exp_f} and \eqref{pert_exp} of amplitudes, the leading part of equation \eqref{sd_m_crit_dt_2} yields
\begin{align}
0&=
\widetilde{\calF}_{1}^{{\rm conn}(0)}(\xi_1) \widetilde{\calF}_{2}^{{\rm conn}(0)}(\xi_1,\xi_2)
+ \pder{\xi_2}
\frac{\widetilde{\calF}_{1}^{{\rm conn}(0)}(\xi_1) - \widetilde{\calF}_{1}^{{\rm conn}(0)}(\xi_2)}{\xi_1 - \xi_2}
\nonumber\\
&\ \ \ \
-\left(\Omega_1(\xi_1)\, 
\tilde{f}_{2}^{{\rm conn}(0)}(\xi_1, \xi_2)\right)_{\mathrm{reg}(\xi_1)}
-\pder{\xi_2} 
\frac{\left(\xi_1^{-1/2}\xi_2^{1/2}\Omega_1(\xi_1)\right)_{\mathrm{reg}(\xi_1)}-\Omega_1(\xi_2)_{\mathrm{reg}(\xi_2)}}
{\xi_1-\xi_2}
\nonumber\\
&\ \ \ \
+\tilde{C}_{1}^{(0)}(\xi_2)
\,,
\label{cyl_eq_dt}
\end{align}
where $\tilde{C}_{1}^{(0)}(\xi_2)$ is a function of $\xi_2$.
Substituting the disk amplitude \eqref{disk_amp_mdt} into equation \eqref{cyl_eq_dt}, we obtain
\begin{align}
0=
\widetilde{\calF}_{2}^{{\rm conn}(0)}(\xi_1,\xi_2) \sqrt{\sigma(\xi_1)}
+\pder{\xi_2}
\frac{\sqrt{\sigma(\xi_1)}-\sqrt{\sigma(\xi_2)}}{\xi_1-\xi_2}
+\frac{R_{2}^{(0)}(\xi_1, \xi_2)}{M(\xi_1)}\,,
\label{cyl_eq_dt2}
\end{align}
where
\begin{align}
R_{2}^{(0)}(\xi_1, \xi_2)&=
\pder{\xi_2}
\frac{\left(M(\xi_1) - M(\xi_2)\right)\sqrt{\sigma(\xi_2)}}{\xi_1 - \xi_2}
- \left(\Omega_1(\xi_1)\, 
\tilde{f}_{2}^{{\rm conn}(0)}(\xi_1, \xi_2)\right)_{\mathrm{reg}(\xi_1)}
\nonumber\\
&\ \ \ \
-\pder{\xi_2} 
\frac{\left(\xi_1^{-1/2}\xi_2^{1/2}\Omega_1(\xi_1)\right)_{\mathrm{reg}(\xi_1)}-\Omega_1(\xi_2)_{\mathrm{reg}(\xi_2)}}
{\xi_1-\xi_2}
+\tilde{C}_{1}^{(0)}(\xi_2)\,,
\end{align}
is a polynomial in $\xi_1$ of degree $m-2$.
Assuming that, as a function of $\xi_1$, 
$\widetilde{\calF}_{2}^{{\rm conn}(0)}(\xi_1,\xi_2)$ has no poles at the zeros of 
$M(\xi_1)$, a polynomial in $\xi_1$ of degree $m-1$,  
we see that the function $R_{2}^{(0)}(\xi_1, \xi_2)$ must vanish. 
As a result, the cylinder amplitude is given by \cite{Eynard:2004mh}:
\begin{align}
\widetilde{\calF}_{2}^{{\rm conn}(0)}(\xi_1,\xi_2)&=
-\frac{1}{\sqrt{\sigma(\xi_1)}}\, \pder{\xi_2}
\frac{\sqrt{\sigma(\xi_1)}-\sqrt{\sigma(\xi_2)}}{\xi_1-\xi_2}
=
\frac{1}{2\sqrt{\sigma(\xi_1)\sigma(\xi_2)}
\left(\sqrt{\sigma(\xi_1)}+\sqrt{\sigma(\xi_2)}\right)^2}
\nonumber
\\
&=
\frac{1}{(\xi_1-\xi_2)^2}
\left(\frac{(\xi_1+\xi_2)/2 + \alpha(\bm{\mu})}
{\sqrt{\xi_1+\alpha(\bm{\mu})}\sqrt{\xi_2+\alpha(\bm{\mu})}}
-1\right).
\label{cyl_dt}
\end{align}

\subsubsection*{Topological recursion}

Applying the perturbative expansions \eqref{pert_exp_f} and \eqref{pert_exp} of amplitudes to equation \eqref{sp_sd_m_crit_dt_rev}, 
we obtain, for $2h+N \ge 3$, 
\begin{align}
0&=
\widetilde{\calF}_{N+1}^{{\rm conn}(h-1)}(\xi_1,\bm{\xi}_{I})
+ 
\mathop{\sum_{h_1+h_2=h}}_{I_1 \cup I_2=I \backslash \{1\}}
\widetilde{\calF}_{|I_1|+1}^{{\rm conn}(h_1)}(\xi_1,\bm{\xi}_{I_1})\,
\widetilde{\calF}_{|I_2|+1}^{{\rm conn}(h_2)}(\xi_1,\bm{\xi}_{I_2})
\nonumber\\
&\ \ \ \
+ 2\, \sum_{i=2}^{\NN}
\pder{\xi_i}
\frac{\widetilde{\calF}_{N-1}^{{\rm conn}(h)}(\bm{\xi}_{I \backslash \{i\}})
-\widetilde{\calF}_{N-1}^{{\rm conn}(h)}(\bm{\xi}_{I \backslash \{1\}})}
{\xi_1 - \xi_i}
-2 \left(\Omega_1(\xi_1)\, 
\tilde{f}_{N}^{{\rm conn}(h)}(\bm{\xi}_I)\right)_{\mathrm{reg}(\xi_1)}
\nonumber\\
&\ \ \ \
+\tilde{C}_{N-1}^{(h)}(\bm{\xi}_{I \backslash \{1\}})
\,,
\label{sp_pert_m_crit_dt}
\end{align}
where $\tilde{C}_{N-1}^{(h)}(\bm{\xi}_{I \backslash \{1\}})$ is a function of $\bm{\xi}_{I \backslash \{1\}}$.
We then obtain, for $2h+N \ge 3$, 
\begin{align}
\label{sp_sd_N_con_pert}
\widetilde{\calF}_{N}^{{\rm conn}(h)}(\bm{\xi}_I)&=
\frac{(-1)}{2\widetilde{\calF}_1^{{\rm conn}(0)}(\xi_1)}
\Biggl[
\widetilde{\calF}_{N+1}^{{\rm conn}(h-1)}(\xi_1,\bm{\xi}_I)
\nonumber\\
&\hspace{8em}
+
\mathop{\sum_{h_1+h_2=h}}_{I_1 \cup I_2=I \backslash \{1\}}^{\textrm{no (0,1)}}
\widetilde{\calF}_{|I_1|+1}^{{\rm conn}(h_1)}(\xi_1, \bm{\xi}_{I_1})\,
\widetilde{\calF}_{|I_2|+1}^{{\rm conn}(h_2)}(\xi_1, \bm{\xi}_{I_2})
\nonumber\\
&\hspace{8em}
+2 \sum_{i=2}^{N} 
\frac{\widetilde{\calF}_{N-1}^{{\rm conn}(h)}(\bm{\xi}_{I \backslash\{i\}})}{(\xi_1-\xi_i)^2}\Biggr]
+\frac{R_{N}^{(h)}(\bm{\xi}_I)}{\widetilde{\calF}_1^{{\rm conn}(0)}(\xi_1)}\,,
\end{align}
where ``no (0,1)'' on the sum of \eqref{sp_sd_N_con_pert} indicates that the terms with $(h, I_1)=(0, \varnothing)$ and $(0, I \backslash \{1\})$ are excluded, and
\begin{align}
R_{N}^{(h)}(\bm{\xi}_I)=
\sum_{i=2}^{N} \pder{\xi_i} 
\frac{\widetilde{\calF}_{N-1}^{{\rm conn}(h)}(\bm{\xi}_{I \backslash \{1\}})}{\xi_1-\xi_i}
+ \left(\Omega_1(\xi_1)\, 
\tilde{f}_{N}^{{\rm conn}(h)}(\bm{\xi}_I)\right)_{\mathrm{reg}(\xi_1)}
-\frac12\tilde{C}_{N-1}^{(h)}(\bm{\xi}_{I \backslash \{1\}})\,.
\end{align}
Note that, viewed as a function of $\xi_1$, $R_{N}^{(h)}(\bm{\xi}_I)$ has no poles on the branch cut 
$[-\alpha(\bm{\mu}), \infty)$ of the disk amplitude \eqref{disk_amp_mdt}.

For $2h+N \le 2$, namely $(h, N)=(0,1)$ and $(0,2)$, we recover the disk amplitude \eqref{disk_amp_mdt} and the cylinder amplitude \eqref{cyl_dt}. 
For later use, we introduce a variable $\eta \in {\IP}^1$ by
\begin{align}
\xi(\eta)=\eta^2-\alpha(\bm{\mu})\,,
\label{local_eta}
\end{align}
and set $\sqrt{\sigma(\xi(\eta))}=\eta$ as a choice of branch.
We also introduce a bidifferential
\begin{align}
B(\eta_1, \eta_2)=\frac{d\eta_1d\eta_2}{(\eta_1-\eta_2)^2}\,,
\label{berg_diff}
\end{align}
which reproduces the cylinder amplitude \eqref{cyl_dt} in the form
\begin{align}
\widetilde{\calF}_{2}^{{\rm conn}(0)}(\xi(\eta_1),\xi(\eta_2))\, d\xi(\eta_1)d\xi(\eta_2)&=
\frac{2\, d\eta_1d\eta_2}{(\eta_1+\eta_2)^2}
=-2B(\eta_1, -\eta_2)
\,.
\label{cyl_local_dt}
\end{align}

For $2h+N \ge 3$, assuming that the amplitudes $\widetilde{\calF}_{N}^{{\rm conn}(h)}(\bm{\xi}_I)$
have no poles away from the branch cut 
$[-\alpha(\bm{\mu}), \infty)$ of the disk amplitude \eqref{disk_amp_mdt}, 
equation \eqref{sp_sd_N_con_pert} is solved as in 
\cite{Eynard:2004mh,Brini:2010fc} 
(in the present context, see Section 4.2.2 of \cite{FMW2025a} for details). 
We now introduce variables $\eta_i$ via 
$
\xi_i=\xi(\eta_i)=\eta_i^2-\alpha(\bm{\mu})
$
in \eqref{local_eta}, and define
\begin{align}
&
dS_{\eta_0}(\eta_1):=
\frac{d\xi(\eta_1)}{\xi(\eta_1)-\xi(\eta_0)}\,
\sqrt{\frac{\sigma(\xi(\eta_0))}{\sigma(\xi(\eta_1))}}
=
\int^{\eta_0}_{-\eta_0} B(\cdot, \eta_1)
=
\frac{2\eta_0\, d\eta_1}{\eta_1^2-\eta_0^2}
\,,
\end{align}
together with multi-differentials%
\footnote{The prefactor $2$ of the bidifferential arises from the normalization of $\GG$. 
If we adopt the change $\GG \to \GG/2$ as a convention, this factor $2$ is no longer needed.
}
\begin{align}
&
\omega_{2}^{(0)}(\eta_{1}, \eta_{2})=2B(\eta_{1}, \eta_{2})=
\frac{2\, d\eta_1d\eta_2}{(\eta_1-\eta_2)^2}\,,
\nonumber\\
&
\omega_{N}^{(h)}(\eta_1, \ldots, \eta_N)
=\widetilde{\calF}_{N}^{{\rm conn}(h)}(\xi(\eta_1),\ldots,\xi(\eta_N))\, 
d\xi(\eta_{1}) \cdots d\xi(\eta_{N})
\ \ \textrm{for}\ \ (h,N)\ne (0,2)\,.
\label{multidiff_dt}
\end{align}
Then the CEO topological recursion \cite{Chekhov:2006vd,Eynard:2007kz} provides solutions to equation \eqref{sp_sd_N_con_pert}:
\begin{align}
\omega_{N}^{(h)}(\bm{\eta}_I)&=
\mathop{\mathrm{Res}}_{\eta_0=0}\,
K_{\eta_0}(\eta_1)
\Biggl[\omega_{N+1}^{(h-1)}(\eta_0,-\eta_0,\bm{\eta}_{I\backslash \{1\}})
\nonumber\\
&\hspace{8em}
+\mathop{\sum_{h_1+h_2=h}}_{I_1 \cup I_2=I \backslash \{1\}}^{\textrm{no (0,1)}}
\omega_{|I_1|+1}^{(h_1)}(\eta_0, \bm{\eta}_{I_1})\,
\omega_{|I_2|+1}^{(h_2)}(-\eta_0, \bm{\eta}_{I_2})\Biggr]\,,
\label{top_rec_multi}
\end{align}
for the one-cut spectral curve given by 
\begin{align}
\xi=\xi(\eta)=\eta^2-\alpha(\bm{\mu})\,,
\quad
y=\widetilde{\calF}_{1}^{{\rm conn}(0)}(\xi(\eta))=M(\xi(\eta))\, \eta\,,
\label{sp_curve_dt_multi}
\end{align} 
where the recursion kernel
\begin{align}
K_{\eta_0}(\eta_1):=
\frac{dS_{\eta_0}(\eta_1)}{4\omega_1^{(0)}(\eta_0)}
=
\frac{d\eta_1}{4M(\xi(\eta_0))\, \eta_0 \left(\eta_1^2-\eta_0^2\right)d\eta_0}\,,
\end{align}
is defined.

This establishes a direct link between the mode-expanded Hamiltonian formalism of the $m$-th multicritical DT and the CEO topological recursion formalism.
In Appendices \ref{app:pure_dt} and \ref{app:m3_dt}, 
we list several amplitudes obtained from the topological recursion \eqref{top_rec_multi} for the spectral curves \eqref{m2_dt_sp} and \eqref{m3_dt_sp}, corresponding respectively to pure DT and $m=3$ multicritical DT, with specialized choices of the cosmological constants.

\subsubsection{Relation between multicritical DT and $(2,2m-1)$ minimal string}
\label{sec:MultiDTandJTgravity}

The conformal-background description of the $(2,2m-1)$ minimal string is already well established. Its disk amplitude was computed explicitly in \cite{Moore:1991ir}, and agreement with the matrix-model description was confirmed, at least for low values of $m$, in \cite{SFT:GK}. The associated minimal-string spectral curve also has a standard geometric description; in the usual uniformization it is expressed in terms of Chebyshev polynomials \cite{Seiberg:2003nm} (see also \cite{Gregori:2021tvs,Fuji:2023wcx}). More recent work \cite{Gregori:2021tvs} has emphasized the distinction between the generic multicritical background and the conformal minimal-string background and has analyzed the perturbative expansion generated by topological recursion on the latter spectral curve, in agreement with the corresponding string-equation description. Thus, neither the conformal-background minimal-string spectral curve nor the application of topological recursion to this curve is new to the present work.

Starting from the string-field-theoretic Hamiltonian of the $m$-th multicritical DT, we derive the SD equations directly, without passing through a matrix-model realization, and obtain the one-cut spectral curve \eqref{sp_curve_dt_multi} from the one-cut solution described in Section~\ref{sec:TR_DT}.
We then show that the explicit specialization \eqref{MultiDTtoJTgravity} of the Hamiltonian cosmological constants $\mu_i$ reduces the corresponding disk amplitude to that of the $(2,2m-1)$ minimal string in the conformal background.
Under this specialization, the spectral curve associated with this one-cut solution agrees, up to the conventions used here, with the standard minimal-string spectral curve, while the corresponding CEO recursion agrees with the known topological recursion on that curve.
The new information provided here is thus the explicit Hamiltonian-to-minimal-string identification, in particular the specialization of the couplings $\{\mu_i\}$, rather than a new minimal-string spectral curve or a new topological-recursion formalism. As a further consequence, we also show that, in the limit $m\to\infty$, the disk amplitude reproduces the Laplace dual of the JT-gravity disk amplitude.

\begin{prop}\label{prop:sp_jt}
When all cosmological constants $\cc_2, \cc_3, \ldots, \cc_m$ 
in the $m$-th multicritical DT are expressed 
in terms of a single cosmological constant $\cc$ as
\begin{align}\label{MultiDTtoJTgravity}
\cc_{2k}=
-\frac{(2m-1)\, (2m-2k-3)!!}{k!\, (2m-4k-1)!!}\left(-\frac{\mu}{8}\right)^k,
\quad
\cc_{2k+1}=0\,,
\end{align}
for $k \in \Dbl{N}$, 
the disk amplitude, for the one-cut solution \eqref{disk_amp_mdt}, takes the form
\begin{equation}\label{DiskAmplitudeJTgravity}
\widetilde{\calF}_1^{{\rm conn}(0)}{\negtinyspace}(\xi)
\,=\,
\frac{(-1)^{m-1} \cc^{(2m-1)/4}}{2^{m-3/2}}
\sin{\negtrpltinyspace}\bigg(
  \frac{2m{\negdbltinyspace}-{\negtrpltinyspace}1}{2} 
  \arccos\!\Big({\negsxpltinyspace}-{\negdbltinyspace}
  \frac{\xi}{\sqrt{\cc}}\Big)
\!\bigg)
\,.
\end{equation}
\end{prop}

\begin{proof}
We rewrite the disk amplitude \eqref{DiskAmplitudeJTgravity} as
\begin{align}
\widetilde{\calF}_1^{{\rm conn}(0)}{\negtinyspace}(\xi)
&=
\frac{(-1)^{m-1} \cc^{(2m-1)/4}}{2^{m-3/2}}
\sin \biggl(
\left(2m-1 \right) 
\arcsin \biggl(\frac{\sqrt{\xi+\sqrt{\mu}}}{\sqrt{2}\, \cc^{1/4}}\biggr)
\biggr)
\nonumber\\
&=
\frac{\cc^{(2m-1)/4}}{2^{m-3/2}}\, 
T_{2m-1}\biggl(\frac{\sqrt{\xi+\sqrt{\mu}}}{\sqrt{2}\, \cc^{1/4}}\biggr)\,,
\label{disk_jt1}
\end{align}
where $T_n(\cos\theta)=\cos(n\theta)$, with a non-negative integer $n$, is the Chebyshev polynomial of the first kind in $\cos \theta$, 
and we have used, in the second equality, the identity
$T_{2m-1}(x)=(-1)^{m-1}\sin((2m-1)\arcsin x)$, valid for the odd positive integer $n=2m-1$.%
Employing the pseudo-Chebyshev function $T_{r}(\cos\theta)=\cos(r\theta)$ for an integer or a half-integer $r$, which satisfies the relation $T_{rs}(x)=T_{r}(T_{s}(x))$, 
we can further rewrite equation \eqref{disk_jt1} as
\begin{align}
\widetilde{\calF}_1^{{\rm conn}(0)}{\negtinyspace}(\xi)
&=
\frac{\cc^{(2m-1)/4}}{2^{m-3/2}}\, 
T_{m-1/2}\biggl(T_2
\biggl(\frac{\sqrt{\xi+\sqrt{\mu}}}{\sqrt{2}\, \cc^{1/4}}\biggr)\biggr)
=
\frac{\cc^{(2m-1)/4}}{2^{m-3/2}}\, 
T_{m-1/2}\biggl(\frac{\xi}{\sqrt{\cc}}\biggr)\,,
\label{disk_jt2}
\end{align}
where we have used $T_2(x)=2x^2-1$ in the second equality.
With the formula 
$T_s(x)=(x-\sqrt{x^2-1})^s/2 + (x+\sqrt{x^2-1})^s/2$ for the pseudo-Chebyshev function, we expand equation \eqref{disk_jt2} around $\xi=\infty$ as follows:
\begin{align}
\widetilde{\calF}_1^{{\rm conn}(0)}{\negtinyspace}(\xi)
&=
\left(\frac{\xi}{2}\right)^{m-1/2}
\left(\left(1-\sqrt{1-\frac{\mu}{\xi^2}}\right)^{m-1/2}
+\left(1+\sqrt{1-\frac{\mu}{\xi^2}}\right)^{m-1/2}\right)
\nonumber\\
&=
2 \left(\frac{\xi}{2}\right)^{m-1/2}
\sum_{k=0}^{\infty}
\frac{\left(m-\frac12\right)\left(m-\frac32\right)\cdots \left(m-2k+\frac12\right)}{(2k)!}
\left(1-\frac{\mu}{\xi^2}\right)^k
\nonumber\\
&=
2 \left(\frac{\xi}{2}\right)^{m-1/2}
{}_2F_1\Big(-\frac{m}{2}+\frac14,-\frac{m}{2}+\frac34;\frac12;1-\frac{\mu}{\xi^2}\Big)\,,
\label{disk_jt3}
\end{align}
where 
\begin{align}
{}_2F_1(\alpha, \beta;\gamma;z)=
\sum_{k=0}^{\infty} 
\frac{\left(\alpha\right)_k \left(\beta\right)_k}{\left(\gamma\right)_k}\,
\frac{z^k}{k!}\,,
\end{align}
is the Gauss hypergeometric function, and 
$(\nu)_k=\nu(\nu+1)\cdots (\nu+k-1)$ denotes the Pochhammer symbol.
By applying the linear transformation formula
\begin{align}
{}_2F_1(\alpha, \beta;\gamma;z)&=
\frac{\Gamma(\gamma)\Gamma(\gamma-\alpha-\beta)}
{\Gamma(\gamma-\alpha)\Gamma(\gamma-\beta)}\,
{}_2F_1(\alpha, \beta;\alpha+\beta-\gamma+1;1-z)
\nonumber\\
&
+
\frac{\Gamma(\gamma)\Gamma(\alpha+\beta-\gamma)}
{\Gamma(\alpha)\Gamma(\beta)}\, (1-z)^{\gamma-\alpha-\beta}\, 
{}_2F_1(\gamma-\alpha, \gamma-\beta;\gamma-\alpha-\beta+1;1-z)\,,
\nonumber
\end{align}
and the Legendre duplication formula 
$\Gamma(z)\Gamma(z+1/2)=2^{1-2z}\Gamma(1/2)\Gamma(2z)$,
equation \eqref{disk_jt3} yields
\begin{align}
\widetilde{\calF}_1^{{\rm conn}(0)}{\negtinyspace}(\xi)
&=
\xi^{m-1/2}\, 
{}_2F_1\Big(-\frac{m}{2}+\frac14,-\frac{m}{2}+\frac34;-m+\frac32;\frac{\mu}{\xi^2}\Big)
\nonumber\\
&\ \
+
\left(\frac{\sqrt{\mu}}{2}\right)^{2m-1}
\xi^{-m+1/2}\, 
{}_2F_1\Big(\frac{m}{2}-\frac14,\frac{m}{2}+\frac14;m+\frac12;\frac{\mu}{\xi^2}\Big)
\nonumber\\
&=
\xi^{m-1/2} - \sum_{k=1}^{m-1} \mu_{2k}\, \xi^{m-2k-1/2}
+O(\xi^{-m+1/2})\,,
\label{disk_jt4}
\end{align}
for $\xi \to \infty$, 
where $\mu_{2k}$ denotes the cosmological constants given by \eqref{MultiDTtoJTgravity}, i.e.,
\begin{align}
\cc_{2k}=
-\frac{(2m-1)\, (2m-2k-3)!!}{k!\, (2m-4k-1)!!}\left(-\frac{\mu}{8}\right)^k.
\end{align}

As a result, using equation \eqref{disk_jt1}, we see that the specialization \eqref{MultiDTtoJTgravity} of the cosmological constants determines the polynomial 
\begin{align}
M(\xi)&=
\frac{\cc^{(2m-1)/4}}{2^{m-3/2} \sqrt{\xi+\sqrt{\mu}}}\, 
T_{2m-1}\biggl(\frac{\sqrt{\xi+\sqrt{\mu}}}{\sqrt{2}\, \cc^{1/4}}\biggr)
\nonumber\\
&=
\sum_{k=0}^{m-1} 
\frac{\left(2m-1\right) \left(m-1+k\right)!}{\left(m-1-k\right)!\left(2k+1\right)!}
\left(-\frac{\sqrt{\cc}}{2}\right)^{m-1-k}
\left(\xi+\sqrt{\cc}\right)^{k},
\end{align}
in $\xi$ and fixes $\alpha(\bm{\mu})=\mu$ for 
the one-cut disk amplitude \eqref{disk_amp_mdt}.
This proves Proposition \ref{prop:sp_jt}.
\end{proof}

The specialization \eqref{MultiDTtoJTgravity} of cosmological constants yields the so-called
``conformal background'' in the study of two-dimensional quantum gravity
coupled to the $(2,2m-1)$ minimal models \cite{Moore:1991ir,SFT:GK}. As a consequence, the disk
amplitude  \rf{DiskAmplitudeJTgravity} coincides with that of the $(2,2m-1)$ minimal string in the conformal background.

Furthermore, by specializing the cosmological constant $\cc$ such that  
$\cc^{1/4} = (2m-1)/(2^{3/2} \pi)$ and taking the parameter $m\to \infty$, 
the disk amplitude with an appropriate normalization factor yields
\begin{align}
\frac{(-1)^{m-1} 2^{m-3/2}}{2\pi \cc^{(2m-1)/4}}\, 
\widetilde{\calF}_1^{{\rm conn}(0)}{\negtinyspace}(x-\sqrt{\cc})
\ \
\mathop{\longrightarrow}^{m \to \infty}
\ \
\frac{1}{2\pi} \sin(2\pi \sqrt{x})\,,
\label{mirz_sp_curve}
\end{align}
whose Laplace transform reproduces the disk amplitude of JT gravity \cite{Stanford:2017thb,Saad:2019lba}, 
which is a theory of two-dimensional gravity with a dilaton field.%
\footnote{Eynard and Orantin \cite{Eynard:2007fi} showed 
that the CEO topological recursion applied to the spectral curve 
\eqref{mirz_sp_curve} yields the  
Laplace transforms of the Weil-Petersson volumes of the moduli spaces of hyperbolic bordered Riemann surfaces obtained via Mirzakhani's recursion \cite{Mirzakhani:2006fta}.}
Here, we emphasize that, in the limit $m \to \infty$, multicritical DT 
contains countably many cosmological constants $\cc_2$, $\cc_3$, \ldots, 
and therefore this limit does not merely correspond to flat two-dimensional space as the dominant configuration.

\section{Causal Dynamical Triangulations (CDT)}
\label{sec:CDT}

In this section we proceed in a way similar to the previous one: we introduce the $m$-th multicritical CDT, whose Hamiltonian is expressed through the relevant $W_{-2}^{(3)}$ operator in the full $W^{(3)}$ representation, and show that the Hamiltonian-derived SD equations admit a one-cut disk solution whose perturbative amplitudes are generated by the CEO topological recursion.
We begin by reviewing pure CDT, which corresponds to the $m=2$ case.

\subsection{Pure CDT}
\label{sec:PureCDT}

\subsubsection{Background}\label{sec:PureCDT_bg}

Before reviewing the formulation of pure CDT, let us briefly recall the theoretical
background relevant to this case. Pure CDT was introduced as a causal counterpart of DT,
in which the underlying triangulations are restricted so as to admit a distinguished time
direction \cite{CDT:AL}. In later works, a string-field-theoretic Hamiltonian formalism for pure CDT
was constructed \cite{CDT:SFT:ALWWZ}, and it was shown that the corresponding SD equations agree
with those of the associated matrix model \cite{CDT:MM:ALWWZ,CDT:MM:ABW}. Thus, in the pure CDT case, the Hamiltonian
description and the matrix-model description are already known to be compatible at the
level of the SD equations.

At the same time, the mode-expanded Hamiltonian realizations of DT and CDT differ. Whereas the pure-DT Hamiltonian is expressed through the relevant $W_{-2}^{(3)}$ operator in the two-reduced representation, the pure-CDT Hamiltonian uses the corresponding $W_{-2}^{(3)}$ operator in the full $W^{(3)}$ representation. Thus, the distinction relevant here is at the level of the operator realization entering the Hamiltonian.

In this section, 
we summarize the formulation of pure CDT from this viewpoint. 
We first introduce the Hamiltonian and operator formalism, 
which serve as the basic starting point for 
the multicritical generalization discussed in Section~\ref{sec:MultiCDT}, 
while also fixing the notation and conventions 
used throughout the subsequent discussion.

\subsubsection{Formulation}
\label{sec:PureCDTdef}

Pure CDT is equivalent to pure DT,
except that the critical point is different.
According to \cite{CDT:SFT:AW}, 
a continuum theory of pure CDT 
can also be formulated using mode expansions.
We briefly summarize this construction below.

In the continuum limit, the generating function for pure CDT is given by \eqref{GeneratingFunModeExpansion}, with the $T\to\infty$ limit understood in the same formal stationary sense as specified in Section~\ref{sec:PureDTdef}, together with the commutation relations \eqref{CommutationRelationPhi} and the vacuum conditions \eqref{vacuumCondition_phi}.

In this case 
the Hamiltonian $\Hop$ is defined by 
\begin{eqnarray}\label{PureCDT_HamiltonianModeExpansion}
\Hop \!\!&\define&\!\!
   -\, 2 g{\tinyspace}
       \phi_{{\tinyspace}2}
 {\dbltinyspace}+{\dbltinyspace}
       \cc_2{\tinyspace} \phi_1
 {\dbltinyspace}-{\dbltinyspace}
       g{\trehalftinyspace}\GG
       \phi_1^2
\nonumber\\&&\!\!
 -\>
   \sum_{\ell=1}^\infty
     \phi_{\ell+1}^\dagger {\dbltinyspace} \ell \phi_\ell
 \,+\,
   \cc_2
   \sum_{\ell=2}^\infty
     \phi_{\ell-1}^\dagger
     {\dbltinyspace} \ell \phi_\ell
 \,-{\tinyspace}
   2{\halftinyspace}g
   \sum_{\ell=3}^\infty
     \phi_{\ell-2}^\dagger
     {\dbltinyspace} \ell \phi_\ell
\nonumber\\&&\!\!
 -\> g
  \sum_{\ell=4}^\infty {\tinyspace} \sum_{n=1}^{\ell-3}
    \phi_n^\dagger \phi_{\ell-n-2}^\dagger
    {\dbltinyspace} \ell \phi_\ell
\,-{\tinyspace}
  g{\trehalftinyspace}\GG
  \sum_{\ell=1}^\infty {\tinyspace} \sum_{n=\max(3-\ell,1)}^\infty\!\!
    \phi_{n+\ell-2}^\dagger
    {\dbltinyspace} n {\halftinyspace} \phi_n
    {\dbltinyspace} \ell \phi_\ell
\,,
\end{eqnarray}
where 
$\cc_2$ is the cosmological constant and  
$\GG$ is introduced to count its number of handles. 
The Hamiltonian \rf{PureCDT_HamiltonianModeExpansion} 
satisfies the no-big-bang condition \rf{NoBigBangCondition}.

In the case of pure CDT, 
the star operation is defined by \rf{StarOpPsiModeExpansion}, 
while the operators $\alpha_n$ are given by  
\rf{DefAlphaOperatorPureDT} with 
$p {\negdbltinyspace}={\negdbltinyspace} 1$ and 
\begin{equation}\label{CoherentEigenValuesPureCDT}
\nu \define 
\frac{1}{\sqrt{\GG}}
\,,
\quad
\lambda_1 \define 
-\,
\frac{\cc_2}{2g\sqrt{\GG}}
\,,
\quad
\lambda_{{\tinyspace}3} \define 
\frac{1}{6g\sqrt{\GG}}
\,,
\quad
\lambda_{{\tinyspace}n} \define 0
\,,
\quad\mbox{[\,$n {\negdbltinyspace}\in{\negdbltinyspace} \Dbl{N}$, 
              $n {\negtrpltinyspace}\neq{\negtrpltinyspace} 1$, $3$
          \!]}
\,.
\end{equation}
The operators $\a_n$ satisfy \rf{CommutationRelationAlphaFlavor}.
Applying the star operation \rf{StarOpDefModeExpansion} to the Hamiltonian 
\rf{PureCDT_HamiltonianModeExpansion} yields
\begin{equation}\label{pureCDT_HamiltonianModeExpansionStarW}
\Hop
\,=\,
-\,g \sqrt{\GG} {\qdrpltinyspace}
 W^{{\dbltinyspace}(3)*}_{{\negdbltinyspace}-2}
\,+\, Y^*
\,,
\end{equation}
where 
$W^{{\dbltinyspace}(3)}_{{\negdbltinyspace}n}$
[{\tinyspace}$n \!\in\! \Dbl{Z}${\trpltinyspace}]
denotes the $W^{(3)}$ operators defined by 
\begin{equation}\label{W3operator}
W^{{\dbltinyspace}(3)}_{{\negdbltinyspace}n}
\,\define\,
\onethird{\negtrpltinyspace}
\sum_{k+l+m
      {\tinyspace}={\tinyspace}n}
{\negtrpltinyspace}{\negtrpltinyspace}{\negtrpltinyspace}
 : \a_k\a_l\a_m :
\,,
\qquad
\mbox{[\,$n$, $k$, $l$, $m {\negdbltinyspace}\in{\negdbltinyspace} \Dbl{Z}$\,]}
\,,
\end{equation}
and $Y$ is introduced so as to satisfy 
the no-big-bang condition \rf{NoBigBangCondition} and has the form
\begin{equation}\label{PureCDT_Y}
Y
\,=\,
\frac{1}{\sqrt{\GG}}
\Big({\dbltinyspace}
  \frac{\a_4}{4 {\halftinyspace} g}{\tinyspace}
  {\tinyspace}-{\tinyspace}
  \frac{\cc_2 {\tinyspace} \a_2}{2 {\halftinyspace} g}{\tinyspace}
  {\tinyspace}+{\tinyspace}
  \a_1
{\negtinyspace}\Big)
\,.
\end{equation}
Equations \eqref{pureCDT_HamiltonianModeExpansionStarW}--\eqref{PureCDT_Y} specify the sense in which the full $W^{(3)}$ representation enters the CDT formulation: the Hamiltonian is expressed in terms of the particular mode $W_{-2}^{(3)}$ defined in \eqref{W3operator}.

For a causal theory, the choice of whether to impose the no-big-bang condition leads to distinct models.
In \cite{CDT:SFT:AW}, the model was defined by setting $Y=0$, whereas in this paper we adopt the choice \eqref{PureCDT_Y}, which coincides with that of \cite{CDT:SFT:ALWWZ} and ensures that the no-big-bang condition is satisfied. Thus, the no-big-bang condition is part of the definition of the CDT Hamiltonian considered here.

\subsubsection{Amplitudes and Schwinger-Dyson equation}
\label{sec:PureCDTamplitudes}

For the Hamiltonian \eqref{PureCDT_HamiltonianModeExpansion},
the amplitudes $f_\NN(\ell_1,\ldots,\ell_\NN)$ and 
the connected amplitudes $f_\NN^{{\rm conn}}(\ell_1,\ldots,\ell_\NN)$ are 
defined by \eqref{GeneralAmplitudes} and \eqref{GeneralConnectedAmplitudes}, respectively. 
Instead of the Laplace-transformed amplitudes \eqref{laplace_amp}, \eqref{laplace_amp_conn}, and \eqref{fr_energy}, we define
\begin{align}
\tilde{f}_\NN(\xi_1,\ldots,\xi_\NN)&=
\sum_{\ell_1=1}^\infty
\ldots
\sum_{\ell_\NN=1}^\infty
\xi_1^{- \ell_1 - 1}
\ldots
\xi_\NN^{- \ell_\NN - 1}
f_\NN(\ell_1,\ldots,\ell_\NN)
\,,
\label{laplace_amp_cdt}
\\
\tilde{f}_\NN^{{\rm conn}}(\xi_1,\ldots,\xi_\NN)&=
\sum_{\ell_1=1}^\infty
\ldots
\sum_{\ell_\NN=1}^\infty
\xi_1^{- \ell_1 - 1}
\ldots
\xi_\NN^{- \ell_\NN - 1}
f_\NN^{{\rm conn}}(\ell_1,\ldots,\ell_\NN)
\,,
\label{laplace_amp_conn_cdt}
\end{align}
and
\begin{align}
\widetilde{\calF}_\NN^{\rm conn}(\xi_1,\ldots,\xi_\NN)&=
\Omega_1(\xi_1)\, \delta_{N,1}
+
\tilde{f}_\NN^{{\rm conn}}(\xi_1,\ldots,\xi_\NN)
\,,
\label{fr_energy_cdt}
\end{align}
where, for pure CDT, 
\begin{align}
\Omega_1(\xi_1)=
\frac{1}{\xi_1}+\frac{1}{2g}\left(\xi_1^2 - \mu_2\right).
\label{om1_pure_cdt}
\end{align}
Note that the $\Omega_2(\xi_1, \xi_2)$ term in \eqref{fr_energy} for the DT case is not introduced in \eqref{fr_energy_cdt} for the CDT case. 
Furthermore, owing to the no-big-bang condition \rf{NoBigBangCondition}, the same form of the SD equation \eqref{sd_eq} is also obtained.

\subsection{Multicritical CDT}
\label{sec:MultiCDT}

\subsubsection{Background}\label{sec:MultiCDT_bg}

The $m$-th multicritical CDT considered in this paper is introduced as a causal analogue of the $m$-th multicritical DT under the no-big-bang condition, within a common mode-expanded Hamiltonian formalism. Related matrix-model descriptions of multicritical CDT have been studied previously, and our model includes, as special cases, earlier models such as those discussed in \cite{Atkin:2012ka,CDT:AGGS}.

What had not been made explicit, however, was the direct derivation of the SD equations from the Hamiltonian formalism itself, together with the construction of a one-cut disk solution and of the perturbative amplitudes around it by the CEO topological recursion. Here we derive these equations explicitly and construct the corresponding spectral curve.
In this way, the multicritical CDT model provides the causal counterpart of the multicritical DT analysis in Section~\ref{sec:MultiDT}, while at the same time clarifying how the previously studied CDT-type models are incorporated into this common mode-expanded Hamiltonian formalism.

\subsubsection{Formulation}
\label{sec:MultiCDTdef}

We introduce the $m$-th multicritical CDT as an extension of the pure CDT constructed 
in complete analogy with the $m$-th multicritical DT. 
Pure CDT discussed in Section \ref{sec:PureCDT} 
coincides with the $m{\negtrpltinyspace}={\negtrpltinyspace}2$ multicritical CDT.
In the continuum limit, 
the generating function for multicritical CDT 
is given by \rf{GeneratingFunModeExpansion},
together with \rf{CommutationRelationPhi} and \rf{vacuumCondition_phi}.
The only difference from pure CDT lies in the Hamiltonian.
To recover pure CDT as a special case, 
we perform a shift of the operators
$\alpha_n$ (i.e., $\phi_n$ for $n = -1$ and $-3$)
in the $W_{-2}^{(3)}$ operator defined in the full representation.

The Hamiltonian before this shift is 
\begin{eqnarray}\label{OriginalCDT_HamiltonianModeExpansion}
\Hop^{({\rm origin})} 
\!\!&\define&\!\!
-\,g \sqrt{\GG} \,{\dbltinyspace}
 W^{{\dbltinyspace}(3)*}_{{\negdbltinyspace}-2}
\Big|_{\nu=1/\sqrt{\GG},\,\,\lambda_n = 0\,[n\in\dblsmall{N}{\dbltinyspace}]}
\nonumber\\&=&\!\!
   -\, 2{\halftinyspace}g{\tinyspace}
       \phi_{{\tinyspace}2}
 {\dbltinyspace}-{\dbltinyspace}
       g{\trehalftinyspace}\GG{\tinyspace}
       \phi_1^2
 {\dbltinyspace}-{\dbltinyspace}
   2{\halftinyspace}g{\tinyspace}
   \sum_{\ell=3}^\infty
     \phi_{\ell-2}^\dagger
     {\dbltinyspace} \ell \phi_\ell
\nonumber\\&&\!\!
 -\> g
  \sum_{\ell=4}^\infty {\tinyspace} \sum_{n=1}^{\ell-3}
    \phi_n^\dagger \phi_{\ell-n-2}^\dagger
    {\dbltinyspace} \ell \phi_\ell
\,-{\dbltinyspace}
  g{\trehalftinyspace}\GG
  \sum_{\ell=1}^\infty {\tinyspace} \sum_{n=\max(3-\ell,1)}^\infty\!\!
    \phi_{n+\ell-2}^\dagger
    {\dbltinyspace} n {\halftinyspace} \phi_n
    {\dbltinyspace} \ell \phi_\ell
\,.
\end{eqnarray}
This Hamiltonian is equivalent to the 
$W^{{\dbltinyspace}(3)}_{{\negqdrpltinyspace}-2}$
operator \rf{W3operator}. 
The general shift in the $W^{(3)}$ operator is specified by 
\begin{eqnarray}\label{CoherentEigenValuesMultiCDT}
&&\!\!\!
\nu \define 
\frac{1}{\sqrt{\GG}}
\,,
\quad
\lambda_{{\tinyspace}i+1} \define 
-\,
\frac{(-1)^{m-i} \cc_{m-i}}
     {2 g {\tinyspace} (i{\negdbltinyspace}+{\negdbltinyspace}1)
      \sqrt{\GG}}
\quad\mbox{[{\dbltinyspace}%
$i{\negdbltinyspace}={\negdbltinyspace}0$, \ldots, 
$m{\negtrpltinyspace}-{\negtrpltinyspace}2$%
{\dbltinyspace}]}
\,,
\quad
\lambda_{{\tinyspace}m+1} \define 
\frac{1}{2 g {\tinyspace} (m{\negdbltinyspace}+{\negdbltinyspace}1)
         \sqrt{\GG}}
\,,
\nonumber\\
&&\!\!\!
\lambda_{{\tinyspace}n} \define 0
\,,
\quad\mbox{[\,$n {\negdbltinyspace}\in{\negdbltinyspace} \Dbl{N}$, 
              $n {\negtrpltinyspace}\neq{\negtrpltinyspace} 1$, $2$, 
                \ldots,
                $m{\negtrpltinyspace}-{\negtrpltinyspace}2$,
                $m{\negtrpltinyspace}-{\negtrpltinyspace}1$,
                $m{\negtrpltinyspace}+{\negtrpltinyspace}1$
          \!]}
\,.
\end{eqnarray}
The Hamiltonian takes the form 
\rf{pureCDT_HamiltonianModeExpansionStarW} 
under the shift 
\rf{DefAlphaOperatorPureDT} with $p=1$ together with \rf{CoherentEigenValuesMultiCDT},
where 
the operator $Y^*$ is a polynomial in the creation operators 
$\phi_n^\dagger$ chosen such that the Hamiltonian 
$\Hop$ satisfies the no-big-bang condition \rf{NoBigBangCondition}.
We thus obtain 
the $m$-th multicritical CDT defined by the Hamiltonian $\Hop$, 
\begin{eqnarray}\label{MultiCDT_HamiltonianModeExpansion}
\Hop \!\!&=&\!\!
\Hop^{({\rm origin})}%
\Big|_{\phi_\ell {\tinyspace}\to{\tinyspace}
       \phi_\ell {\tinyspace}+{\tinyspace}
       \lambda_\ell \sqrt{1/\GG}}
{\dbltinyspace}+
Y^*
\nonumber\\&=&\!\!
   -\, 2 g{\tinyspace}
       \phi_{{\tinyspace}2}
 {\dbltinyspace}+{\dbltinyspace}
       (-1)^m \cc_m{\tinyspace} \phi_1
 {\dbltinyspace}-{\dbltinyspace}
       g{\trehalftinyspace}\GG
       \phi_1^2
\nonumber\\&&
 -\>
   \sum_{\ell=1}^\infty
     \phi_{\ell+m-1}^\dagger {\dbltinyspace} \ell \phi_\ell
 \,+\,
   \sum_{i=2}^{m} {\tinyspace} \sum_{\ell=\max(i-m+2,1)}^\infty\!\!
     (-1)^i \cc_i {\tinyspace} \phi_{\ell+m-i-1}^\dagger
     {\dbltinyspace} \ell \phi_\ell
 \,-\,
   2{\halftinyspace}g
   \sum_{\ell=3}^\infty
     \phi_{\ell-2}^\dagger
     {\dbltinyspace} \ell \phi_\ell
\nonumber\\&&
 -\> g
  \sum_{\ell=4}^\infty {\tinyspace} \sum_{n=1}^{\ell-3}
    \phi_n^\dagger \phi_{\ell-n-2}^\dagger
    {\dbltinyspace} \ell \phi_\ell
\,-\,
  g{\trehalftinyspace}\GG
  \sum_{\ell=1}^\infty {\tinyspace} \sum_{n=\max(3-\ell,1)}^\infty\!\!
    \phi_{n+\ell-2}^\dagger
    {\dbltinyspace} n {\halftinyspace} \phi_n
    {\dbltinyspace} \ell \phi_\ell
\,,
\end{eqnarray}
where 
$\cc_i$ 
[{\tinyspace}$i{\negtrpltinyspace}={\negtrpltinyspace}2$,
 \ldots, $m${\tinyspace}]
are cosmological constants with mass dimension $i$, 
and $\GG$ counts the number of handles of a 2D surface as before.

\subsubsection{Amplitudes and Schwinger-Dyson equation}
\label{sec:MultiCDTamplitudes}

For the Hamiltonian $H$ in \eqref{MultiCDT_HamiltonianModeExpansion} of the $m$-th multicritical CDT, the amplitudes and connected amplitudes are defined by \eqref{GeneralAmplitudes} and \eqref{GeneralConnectedAmplitudes}, respectively. In this subsection, we derive the corresponding SD equations directly from the Hamiltonian under the no-big-bang condition, following the argument of Section~\ref{sec:MultiDTamplitudes}. These equations provide the starting point for the CEO-recursive reformulation developed in the next subsection.

For the Hamiltonian $\Hop^{({\rm origin})}$ in \eqref{OriginalCDT_HamiltonianModeExpansion}, we have
\begin{align}
\begin{split}
\left[\Hop^{({\rm origin})}, \phi_{\ell}^{\dagger}\right]&=
-2g \delta_{\ell,1}\, \GG \phi_1
-2g \delta_{\ell,2} 
- 2g \theta_{\ell, 3}\, \ell\, \phi_{\ell-2}^{\dagger}
\\
&\ \ \ \
-2g\, \ell\, \sum_{n=1}^{\ell-3} \phi_n^{\dagger} \phi_{\ell-n-2}^{\dagger}
-2g\, \ell\, \GG \sum_{n= \mathrm{max}(3-\ell,1)}^{\infty} n\, \phi_{n+\ell-2}^{\dagger} \phi_n\,,
\label{com_H_1_cdt}
\end{split}
\end{align}
and hence one finds
\begin{align}
&
-\frac{1}{g}\, \phi_{\ell_1}^{\dagger}\cdots \phi_{\ell_{i-1}}^{\dagger}
\left[\Hop^{({\rm origin})}, \phi_{\ell_i}^{\dagger}\right]
\phi_{\ell_{i+1}}^{\dagger}\cdots \phi_{\ell_N}^{\dagger}
\nonumber\\
&=
\phi_{\ell_1}^{\dagger}\cdots \breve{\phi}_{\ell_i}^{\dagger}\cdots
\phi_{\ell_N}^{\dagger}
\left(2 \delta_{\ell_i,1}\, \GG \phi_1
+ 2 \delta_{\ell_i,2} 
+ 2 \theta_{\ell_i, 3}\, \ell_i\, \phi_{\ell_i-2}^{\dagger}
\right)
\nonumber\\
&\ \
+\phi_{\ell_1}^{\dagger}\cdots \breve{\phi}_{\ell_i}^{\dagger}\cdots
\phi_{\ell_N}^{\dagger}\left(
\ell_i \sum_{n=1}^{\ell_i-3} \phi_n^{\dagger} \phi_{\ell_i-n-2}^{\dagger}
+ 2\GG\, \ell_i 
\sum_{n= \mathrm{max}(3-\ell_i,1)}^{\infty} n\, \phi_{n+\ell_i-2}^{\dagger} \phi_n
\right)
\nonumber\\
&\ \
+ 2 \GG \sum_{j=i+1}^N \delta_{\ell_i+\ell_j,2}\, 
\phi_{\ell_1}^{\dagger}\cdots \breve{\phi}_{\ell_i}^{\dagger} \cdots
\breve{\phi}_{\ell_j}^{\dagger} \cdots \phi_{\ell_N}^{\dagger}
\nonumber\\
&\ \
+ 2 \GG \sum_{j=i+1}^N
\ell_i\, \ell_j\, \theta_{\ell_i+\ell_j-2,1}\,
\phi_{\ell_i+\ell_j-2}^{\dagger}\,
\phi_{\ell_1}^{\dagger}\cdots \breve{\phi}_{\ell_i}^{\dagger}
\cdots \breve{\phi}_{\ell_j}^{\dagger} \cdots \phi_{\ell_N}^{\dagger}\,,
\label{com_H_phi_cdt}
\end{align}
where $\breve{\phi}^{\dagger}$ indicates that the corresponding operator $\phi^{\dagger}$ is omitted, and
$\theta_{k,\ell}=1$ if $k\ge \ell$ and $\theta_{k,\ell}=0$ otherwise.
We then find that the Laplace-transformed amplitudes \eqref{laplace_amp_cdt} 
associated with the Hamiltonian \eqref{MultiCDT_HamiltonianModeExpansion} of the $m$-th multicritical CDT satisfy
\begin{align}
0&=
\sum_{i=1}^N \Biggl[\pder{\xi_i}
\tilde{f}_{N+1}(\xi_i, \bm{\xi}_I)
+ 
\frac{1}{g} \pder{\xi_i}
\left(
\xi_i^{m} \tilde{f}_{N}(\bm{\xi}_I)-
\left(\xi_i^{m} \tilde{f}_{N}(\bm{\xi}_I)\right)_{\mathrm{reg}(\xi_i)}
\right)
\nonumber\\
&\hspace{3em}
-\frac{1}{g}
\sum_{p = 0}^{m-2} (-1)^{m-p}\mu_{m-p} \pder{\xi_i}
\left(
\xi_i^{p} \tilde{f}_{N}(\bm{\xi}_I)-
\left(\xi_i^{p} \tilde{f}_{N}(\bm{\xi}_I)\right)_{\mathrm{reg}(\xi_i)}
\right)
\nonumber\\
&\hspace{3em}
+ 2 \pder{\xi_i} \left(\xi_i^{-1} \tilde{f}_{N}(\bm{\xi}_I)\right)
+\GG \sum_{\substack{j=1 \\ (j \ne i)}}^{N} 
\frac{\partial^2}{\partial \xi_i \partial \xi_j}
\frac{\tilde{f}_{N-1}(\bm{\xi}_{I \backslash \{j\}})
-\tilde{f}_{N-1}(\bm{\xi}_{I \backslash \{i\}})}
{\xi_i-\xi_j}
\nonumber\\
&\hspace{3em}
- \left(2 \xi_i^{-3}-(-1)^{m}\frac{\mu_m}{g} \xi_i^{-2}\right)
\tilde{f}_{N-1}(\bm{\xi}_{I \backslash \{i\}})
- \GG \sum_{\substack{j=1 \\ (j \ne i)}}^{N} 
\xi_i^{-2}\xi_j^{-2}\,
\tilde{f}_{N-2}(\bm{\xi}_{I \backslash \{i, j\}})\Biggr]\,.
\label{sd_multi_cdt}
\end{align}

From equation \eqref{sd_multi_cdt}, we deduce a separated SD equation focused on the variable $\xi_1$:
\begin{align}
0&=
\tilde{f}_{N+1}(\xi_1, \bm{\xi}_I)
+ 
2 \left(
\Omega_1(\xi_1)\, \tilde{f}_{N}(\bm{\xi}_I)-
\left(\Omega_1(\xi_1)\,  \tilde{f}_{N}(\bm{\xi}_I)\right)_{\mathrm{reg}(\xi_1)}
\right)
\nonumber\\
&\ \
+\GG \sum_{i=2}^{N} \pder{\xi_i}
\frac{\tilde{f}_{N-1}(\bm{\xi}_{I \backslash \{i\}})
-\tilde{f}_{N-1}(\bm{\xi}_{I \backslash \{1\}})}
{\xi_1-\xi_i}
+ \left(\xi_1^{-2}-(-1)^{m}\frac{\mu_m}{g} \xi_1^{-1}\right)
\tilde{f}_{N-1}(\bm{\xi}_{I \backslash \{1\}})
\nonumber\\
&\ \
+ \GG \sum_{i=2}^{N} 
\xi_1^{-1}\xi_i^{-2}\,
\tilde{f}_{N-2}(\bm{\xi}_{I \backslash \{1, i\}})
+ C_{N-1}(\bm{\xi}_{I \backslash \{1\}})
\,,
\label{sp_sd_multi_cdt}
\end{align}
where $C_{N-1}(\bm{\xi}_{I \backslash \{1\}})$ is a function of $\bm{\xi}_{I \backslash \{1\}}$. Instead of \eqref{om1_pure_cdt}, we introduce
\begin{align}
\Omega_1(\xi)=
\frac{1}{\xi}+
\frac{1}{2g}\left(\xi^m
- \sum_{p = 0}^{m-2} (-1)^{m-p} \mu_{m-p}\, \xi^p
\right).
\label{om1_multi_cdt}
\end{align}
By defining the amplitudes $\widetilde{\calF}_\NN^{\rm conn}(\xi_1,\ldots,\xi_\NN)$ via \eqref{fr_energy_cdt}, 
together with the shift \eqref{om1_multi_cdt} which generalizes \eqref{om1_pure_cdt}, 
we obtain the CDT analogue of Proposition \ref{prop:sd_multi_dt} for multicritical DT (see Appendix \ref{app:proof_prop_cdt} for a proof).

\begin{prop}\label{prop:sd_multi_cdt}
The amplitudes $\widetilde{\calF}_\NN^{\rm conn}(\xi_1,\ldots,\xi_\NN)$ 
determined by equation \eqref{sp_sd_multi_cdt},
which also provide solutions to the SD equation \eqref{sd_multi_cdt}, 
satisfy
\begin{align}
0&=
\widetilde{\calF}_{N+1}^{\rm conn}(\xi_1,\bm{\xi}_{I})
+ 
\sum_{I_1 \cup I_2=I \backslash \{1\}}
\widetilde{\calF}_{|I_1|+1}^{\rm conn}(\xi_1,\bm{\xi}_{I_1})\,
\widetilde{\calF}_{|I_2|+1}^{\rm conn}(\xi_1,\bm{\xi}_{I_2})
\nonumber\\
&\ \ \ \
+ \GG {\negdbltinyspace}
\sum_{i=2}^{\NN}
\pder{\xi_i}
\frac{\widetilde{\calF}_{N-1}^{\rm conn}(\bm{\xi}_{I \backslash \{i\}})
-\widetilde{\calF}_{N-1}^{\rm conn}(\bm{\xi}_{I \backslash \{1\}})}
{\xi_1 - \xi_i}
-2 \left(\Omega_1(\xi_1)\, 
\tilde{f}_{N}^{{\rm conn}}(\bm{\xi}_I)\right)_{\mathrm{reg}(\xi_1)}
\nonumber\\
&\ \ \ \
-\left(\Omega_1(\xi_1)^2\right)_{\mathrm{reg}(\xi_1)}\, \delta_{N,1}
- \GG \pder{\xi_2} 
\frac{\Omega_1(\xi_1)_{\mathrm{reg}(\xi_1)}-\Omega_1(\xi_2)_{\mathrm{reg}(\xi_2)}}
{\xi_1-\xi_2}\, \delta_{N,2}
+\tilde{C}_{N-1}(\bm{\xi}_{I \backslash \{1\}})
\,,
\label{sp_sd_m_crit_cdt_rev}
\end{align}
where $\tilde{C}_{N-1}(\bm{\xi}_{I \backslash \{1\}})$ is a function of $\bm{\xi}_{I \backslash \{1\}}$.
\end{prop}

We remark that equation \eqref{sp_sd_m_crit_cdt_rev} agrees with 
the SD equation (i.e., loop equation) in the Hermitian matrix model \cite{Eynard:2004mh} with potential $V(\phi)$
for the matrix $\phi$ \cite{Atkin:2012ka}
(see also \cite{CDT:MM:ALWWZ} for the pure CDT, \cite{CDT:AGGS} for the $m=3$ multicritical CDT):
\begin{align}
V(\phi)=
-\frac{1}{2g}\left(
\frac{1}{m+1}\, \phi^{m+1}
+
\sum_{p = 1}^{m-1} \frac{(-1)^{m-p} \mu_{m-p+1}}{p}\, \phi^p\right),
\end{align}
under the identification
\begin{align}
\Omega_1(\xi)=
\frac{1}{\xi} - V'(\xi)\,.
\end{align}

\subsubsection{Perturbative amplitudes and topological recursion}
\label{sec:TR_CDT}

As in the DT case discussed in Section \ref{sec:TR_DT}, we consider 
perturbative solutions of equation \eqref{sp_sd_m_crit_cdt_rev} 
around $\GG=0$ by expanding the amplitudes as in \eqref{pert_exp_f} and \eqref{pert_exp}.

\subsubsection*{Disk amplitude}

For $N=1$, equation \eqref{sp_sd_m_crit_cdt_rev}  reads
\begin{align}
0&=
\widetilde{\calF}_{2}^{\rm conn}(\xi,\xi)
+ 
\widetilde{\calF}_{1}^{\rm conn}(\xi)^2
-2 \left(\Omega_1(\xi)\, 
\tilde{f}_{1}^{{\rm conn}}(\xi)\right)_{\mathrm{reg}(\xi)}
-\left(\Omega_1(\xi)^2\right)_{\mathrm{reg}(\xi)}
+\tilde{C}_{0}
\,,
\label{sp_sd_m_crit_cdt_1}
\end{align}
where we note that this has exactly the same form as equation \eqref{sd_m_crit_dt_1}, except for the specific form of $\Omega_1(\xi)$.
Using the perturbative expansions \eqref{pert_exp_f} and \eqref{pert_exp} of the amplitudes, we obtain
\begin{align}
\widetilde{\calF}_{1}^{{\rm conn}(0)}(\xi)^2=
\left(\Omega_1(\xi)^2\right)_{\mathrm{reg}(\xi)}
+2 \left(\Omega_1(\xi)\, 
\tilde{f}_{1}^{{\rm conn}(0)}(\xi)\right)_{\mathrm{reg}(\xi)}
+\tilde{C}_{0}
\,,
\label{disk_eq_mdt_cdt}
\end{align}
where $\tilde{C}_{0}^{(0)}$ is a function of the cosmological constants $\mu_p$ and $g$ that does not depend on $\xi$.

As in the DT case, although equation \eqref{disk_eq_mdt_cdt} does not uniquely exclude other possible analytic structures, the Hamiltonian-derived disk SD equation admits a one-cut solution. 
We focus on this one-cut solution as the natural multicritical analogue of the physical continuum branch selected in the pure CDT case \cite{CDT:MM:ALWWZ,Atkin:2012ka,CDT:AGGS}. 
It takes the form
\begin{align}
\widetilde{\calF}_{1}^{{\rm conn}(0)}(\xi)=
\frac{1}{2g}\left(\xi^{m-1}+\sum_{p=1}^{m-1}t_{m-p}(\bm{\mu})\, \xi^{p-1} \right)
\sqrt{\left(\xi + \alpha(\bm{\mu})\right)\left(\xi + \beta(\bm{\mu})\right)}
=:
M(\xi)\, \sqrt{\sigma(\xi)}\,,
\label{disk_amp_m_cdt}
\end{align}
where we have introduced the notations
\begin{align}
M(\xi)=\frac{1}{2g}\left(\xi^{m-1}+\sum_{p=1}^{m-1}t_{m-p}(\bm{\mu})\, \xi^{p-1} \right),
\quad
\sigma(\xi)=\left(\xi + \alpha(\bm{\mu})\right)\left(\xi + \beta(\bm{\mu})\right).
\end{align}
Substituting this one-cut solution into equation \eqref{disk_eq_mdt_cdt} yields the equations below.
Consequently, the one-cut data $\alpha$, $\beta$, $t_1,\ldots,t_{m-1}$ are determined by the Hamiltonian parameters $\bm{\mu}$ and $g$ through these algebraic relations, rather than being introduced as independent parameters.

Here the $m+1$ functions $\alpha=\alpha(\bm{\mu})$, $\beta=\beta(\bm{\mu})$, $t_1=t_1(\bm{\mu}), \ldots, t_{m-1}=t_{m-1}(\bm{\mu})$, 
which depend on 
$\bm{\mu}=\{\mu_2, \ldots, \mu_{m}\}$, are determined by the $m+1$ equations
\begin{align}
&
\sum_{\substack{a,b=0 \\ (a+b=p)}}^{m-1} t_a t_b 
+ \left(\alpha+\beta\right) \sum_{\substack{a,b=0 \\ (a+b=p-1)}}^{m-1} t_a t_b
+ \alpha \beta \sum_{\substack{a,b=0 \\ (a+b=p-2)}}^{m-1} t_a t_b
\nonumber\\
&=
(-1)^p\left(
\sum_{\substack{a,b=2 \\ (a+b=p)}}^{m} \mu_a \mu_b - 2\mu_p \right)
+4g\, \delta_{p,m+1}\,,
\quad
p=1, \ldots, m+1\,,
\label{sp_det_eq_cdt}
\end{align}
where $t_0 \equiv 1$ and $\mu_1 = \mu_{m+1} \equiv 0$.

For example, when $m=2$, corresponding to the pure CDT, equations \eqref{sp_det_eq_cdt} yield
$$
\alpha+\beta=-2t_1\,,\quad
\alpha \beta = 3t_1^2-2\mu_2\,,\quad
t_1^3 - \mu_2\, t_1 - g = 0\,.
$$
For $m=2$, equations \eqref{sp_det_eq_cdt} admit several algebraic branches.
We identify the physical branch used in the perturbative expansion around $g=0$ by requiring that the disk amplitude have no poles in the region $\mathrm{Re}(\xi)>0$, consistently with the physical branch used in the matrix-model analysis \cite{Atkin:2012ka}.
This condition selects the physical branch among the one-cut solutions admitted by the Hamiltonian-derived SD equation.
With this choice, we obtain $t_1$ expanded around $g=0$ as
\begin{align}
t_1=
-\sqrt{\mu_2} + \frac{g}{2\mu_2} + \frac{3g^2}{8\mu_2^{5/2}}
+ \frac{g^3}{2\mu_2^4} + \frac{105g^4}{128\mu_2^{11/2}}
+ \frac{3g^5}{2\mu_2^7} + \frac{3003g^6}{1024\mu_2^{17/2}}
+ O(g^7)
\,.
\label{pure_cdt_s0_sol}
\end{align} 
From this solution, we obtain (higher-order terms of $g$ are provided in Appendix \ref{app:pure_cdt}),
\begin{align}
\widetilde{\calF}_{1}^{{\rm conn}(0)}(\xi)&=
\frac{\xi^2 -\mu_2}{2g} 
+ \frac{1}{\xi + \sqrt{\mu_2}}
+ \frac{\xi+3\sqrt{\mu_2}}{4\mu_2 (\xi + \sqrt{\mu_2})^3}\, g
+O(g^2)\,.
\label{pure_cdt_disk}
\end{align}

When $m=3$, equations \eqref{sp_det_eq_cdt} yield
\begin{align}
&
\alpha+\beta=-2t_1\,,\quad
\alpha \beta = 3t_1^2 - 2t_2 - 2\mu_2\,,
\nonumber\\
&
2t_{1}^{3}-3t_{1}t_{2}-2\mu_2t_{1}-\mu_3=0\,,\quad
3t_{1}^{4}-2\mu_2t_{1}^{2}-3t_{2}^{2}-4\mu_2t_{2}-\mu_2^{2}-4g=0\,.
\nonumber
\end{align}
By considering the special case $\mu_2=0$ and taking solutions expanded around $g=0$ as
\begin{align}
t_1&=
-\mu_{3}^{{1}/{3}}-\frac{2 g}{3 \mu_{3}}+\frac{2 g^{2}}{3 \mu_{3}^{{7}/{3}}}-\frac{76 g^{3}}{81 \mu_{3}^{{11}/{3}}}+\frac{110 g^{4}}{81 \mu_{3}^{5}}-\frac{412 g^{5}}{243 \mu_{3}^{{19}/{3}}}+\frac{6748 g^{6}}{6561 \mu_{3}^{{23}/{3}}}
+O(g^7)\,,
\nonumber\\
t_2&=
\mu_{3}^{{2}/{3}}+\frac{2 g}{3 \mu_{3}^{{2}/{3}}}-\frac{2 g^{2}}{9 \mu_{3}^{2}}-\frac{4 g^{3}}{81 \mu_{3}^{{10}/{3}}}+\frac{170 g^{4}}{243 \mu_{3}^{{14}/{3}}}-\frac{580 g^{5}}{243 \mu_{3}^{6}}+\frac{42476 g^{6}}{6561 \mu_{3}^{{22}/{3}}}
+O(g^7)\,,
\label{m3_cdt_s0_sol}
\end{align}
we obtain (higher-order terms of $g$ are provided in Appendix \ref{app:m3_cdt}),
\begin{align}
\widetilde{\calF}_{1}^{{\rm conn}(0)}(\xi)&=
\frac{\xi^{3}+\mu_{3}}{2g} 
+ \frac{1}{\xi+\mu_{3}^{{1}/{3}}}
- \frac{\xi + 2 \mu_{3}^{{1}/{3}}}{3\mu_{3} \left(\xi+\mu_{3}^{{1}/{3}}\right)^{3}}\, g
+O(g^2)\,.
\label{m3_cdt_disk}
\end{align}

Here we remark that, in general, 
the one-cut disk amplitude of the $m$-th multicritical CDT with 
$\mu_2=\cdots=\mu_{m-1}=0$ and $\mu_m \ne 0$, 
which has no poles in the region $\mathrm{Re}(\xi)>0$, takes the form \cite{Atkin:2012ka},
\begin{align}
\widetilde{\calF}_{1}^{{\rm conn}(0)}(\xi)&=
\frac{\xi^{m}-(-1)^m \mu_{m}}{2g} 
+ \frac{1}{\xi+\mu_{m}^{{1}/{m}}}
+O(g)\,,
\label{mult_cdt_disk}
\end{align}
around $g=0$.
This follows from equation \eqref{disk_eq_mdt_cdt} using the one-cut solution \eqref{disk_amp_m_cdt} by expanding it around $g=0$ and $\xi=\infty$:
\begin{align*}
\widetilde{\calF}_{1}^{{\rm conn}(0)}(\xi)^2&=
\left(\frac{\xi^{m}-(-1)^m \mu_{m}}{2g}\right)^2 + 
\frac{1}{g}\left( \xi^{m-1} + O(\xi^{m-2})\right) + O(g^0)
\\
&=
\left(
\frac{1}{2g} \sum_{p=0}^{m-1}(- \mu_m^{1/m})^p \xi^{m-1-p} + O(g^0)\right)^2
\left(
\left(\xi + \mu_m^{1/m}\right)^2 + O(g)
\right),
\end{align*}
where the second equality uniquely fixes the subleading terms in $g$ that appear on the right-hand side of the first equality, and thus \eqref{mult_cdt_disk} is obtained.

\subsubsection*{Cylinder amplitude}

For $N=2$, equation \eqref{sp_sd_m_crit_cdt_rev} takes the form
\begin{align}
0&=
\widetilde{\calF}_{3}^{\rm conn}(\xi_1,\xi_1, \xi_2)
+ 
2 \widetilde{\calF}_{1}^{\rm conn}(\xi_1)\, \widetilde{\calF}_{2}^{\rm conn}(\xi_1,\xi_2)
+ \GG {\negdbltinyspace}
\pder{\xi_2}
\frac{\widetilde{\calF}_{1}^{\rm conn}(\xi_1)-\widetilde{\calF}_{1}^{\rm conn}(\xi_2)}
{\xi_1 - \xi_2}
\nonumber\\
&\ \ \ \
-2 \left(\Omega_1(\xi_1)\, 
\tilde{f}_{2}^{{\rm conn}}(\xi_1, \xi_2)\right)_{\mathrm{reg}(\xi_1)}
- \GG \pder{\xi_2}
\frac{\Omega_1(\xi_1)_{\mathrm{reg}(\xi_1)}-\Omega_1(\xi_2)_{\mathrm{reg}(\xi_2)}}
{\xi_1-\xi_2}
+\tilde{C}_{1}(\xi_2)
\,,
\label{sp_sd_m_crit_cdt_2}
\end{align}
where we note that this equation has a form similar to equation \eqref{sd_m_crit_dt_2}, but with the replacement $\GG \to \GG/2$.
We now consider the perturbative expansions \eqref{pert_exp_f} and \eqref{pert_exp} of the amplitudes and adopt the solution \eqref{disk_amp_m_cdt}. 
The leading part of equation \eqref{sp_sd_m_crit_cdt_2} is solved as in \eqref{cyl_dt}, under the assumption that 
$\widetilde{\calF}_{2}^{{\rm conn}(0)}(\xi_1,\xi_2)$, regarded as a function of $\xi_1$, 
has no poles at the zeros of $M(\xi_1)$, 
and we obtain the following expression \cite{Eynard:2004mh}:
\begingroup\makeatletter\def\f@size{11.3}\check@mathfonts
\def\maketag@@@#1{\hbox{\m@th\normalsize\normalfont#1}}%
\begin{align}
\widetilde{\calF}_{2}^{{\rm conn}(0)}(\xi_1,\xi_2)&=
-\frac{1}{2\sqrt{\sigma(\xi_1)}}\, \pder{\xi_2}
\frac{\sqrt{\sigma(\xi_1)}-\sqrt{\sigma(\xi_2)}}{\xi_1-\xi_2}
\nonumber
\\
&=
\frac{1}{2(\xi_1-\xi_2)^2}
\left(\frac{\xi_1\xi_2+\left(\alpha(\bm{\mu})+\beta(\bm{\mu})\right)
\left(\xi_1+\xi_2\right)/2 + \alpha(\bm{\mu})\beta(\bm{\mu})}
{\sqrt{\left(\xi_1+\alpha(\bm{\mu})\right)\left(\xi_1+\beta(\bm{\mu})\right)}
\sqrt{\left(\xi_2+\alpha(\bm{\mu})\right)\left(\xi_2+\beta(\bm{\mu})\right)}}
-1\right).
\label{cyl_cdt}
\end{align}
\endgroup

For example, for pure CDT and 
$m=3$ multicritical CDT with $\mu_2=0$, by adopting the solutions \eqref{pure_cdt_s0_sol} and \eqref{m3_cdt_s0_sol} expanded around $g=0$, 
we obtain, respectively
\begin{align}
\widetilde{\calF}_{2}^{{\tinyspace}{\rm conn}(0)}{\negdbltinyspace}(\xi_1, \xi_2)
&=
\frac{g}{2 \sqrt{\mu_{2}}\, \left(\xi_{1}+\sqrt{\mu_{2}}\right)^{2} \left(\xi_{2}+\sqrt{\mu_{2}}\right)^{2}}
+ O(g^2)\,,
\label{pure_cdt_cyl}
\\
\widetilde{\calF}_{2}^{{\tinyspace}{\rm conn}(0)}{\negdbltinyspace}(\xi_1, \xi_2)
&=
-\frac{g}{3 \mu_{3}^{{2}/{3}} \left(\xi_{1}+\mu_{3}^{{1}/{3}}\right)^{2} \left(\xi_{2}+\mu_{3}^{{1}/{3}}\right)^{2}}
+ O(g^2)\,,
\label{m3_cdt_cyl}
\end{align}
where higher-order terms of $g$ are provided in Appendices \ref{app:pure_cdt} and \ref{app:m3_cdt}.

\subsubsection*{Topological recursion}

By applying the perturbative expansions \eqref{pert_exp_f} and \eqref{pert_exp} of amplitudes to equation \eqref{sp_sd_m_crit_cdt_rev}, 
we obtain, for $2h+N \ge 3$, 
\begin{align}
0&=
\widetilde{\calF}_{N+1}^{{\rm conn}(h-1)}(\xi_1,\bm{\xi}_{I})
+ 
\mathop{\sum_{h_1+h_2=h}}_{I_1 \cup I_2=I \backslash \{1\}}
\widetilde{\calF}_{|I_1|+1}^{{\rm conn}(h_1)}(\xi_1,\bm{\xi}_{I_1})\,
\widetilde{\calF}_{|I_2|+1}^{{\rm conn}(h_2)}(\xi_1,\bm{\xi}_{I_2})
\nonumber\\
&\ \ \ \
+ \sum_{i=2}^{\NN}
\pder{\xi_i}
\frac{\widetilde{\calF}_{N-1}^{{\rm conn}(h)}(\bm{\xi}_{I \backslash \{i\}})
-\widetilde{\calF}_{N-1}^{{\rm conn}(h)}(\bm{\xi}_{I \backslash \{1\}})}
{\xi_1 - \xi_i}
-2 \left(\Omega_1(\xi_1)\, 
\tilde{f}_{N}^{{\rm conn}(h)}(\bm{\xi}_I)\right)_{\mathrm{reg}(\xi_1)}
\nonumber\\
&\ \ \ \
+\tilde{C}_{N-1}^{(h)}(\bm{\xi}_{I \backslash \{1\}})
\,,
\label{sp_pert_m_crit_cdt}
\end{align}
where $\tilde{C}_{N-1}^{(h)}(\bm{\xi}_{I \backslash \{1\}})$ is a function of $\bm{\xi}_{I \backslash \{1\}}$.
This is almost the same form as equation \eqref{sp_pert_m_crit_dt},  
except for the factor $2$ in the first term of the second line.
As in the argument below \eqref{sp_pert_m_crit_dt}, 
we assume that the amplitudes $\widetilde{\calF}_{N}^{{\rm conn}(h)}(\bm{\xi}_I)$ 
with $2h+N \ge 3$ have no poles away from the branch cut of the disk amplitude \eqref{disk_amp_m_cdt}. 
Then, it is straightforward to show that the CEO topological recursion 
\begin{align}
\omega_{N}^{(h)}(\bm{\eta}_I)&=
\sum_{s=\pm 1}
\mathop{\mathrm{Res}}_{\eta_0=s}\,
K_{\eta_0}(\eta_1)
\Biggl[\omega_{N+1}^{(h-1)}(\eta_0,\eta_0^{-1},\bm{\eta}_{I \backslash \{1\}})
\nonumber\\
&\hspace{8em}
+\mathop{\sum_{h_1+h_2=h}}_{I_1 \cup I_2=I \backslash \{1\}}^{\textrm{no (0,1)}}
\omega_{|I_1|+1}^{(h_1)}(\eta_0, \bm{\eta}_{I_1})\,
\omega_{|I_2|+1}^{(h_2)}(\eta_0^{-1}, \bm{\eta}_{I_2})\Biggr]\,,
\label{top_rec_multi_cdt}
\end{align}
solves equation \eqref{sp_pert_m_crit_cdt}.
Here we introduce the Zhukovsky variable $\eta \in {\IP}^1$, by
\begin{align}
\xi(\eta)=-\frac{\alpha(\bm{\mu})+\beta(\bm{\mu})}{2}
+\frac{\alpha(\bm{\mu})-\beta(\bm{\mu})}{4}\left(\eta+\eta^{-1}\right),
\end{align}
which provides the spectral curve
\begin{align}
\xi=\xi(\eta)\,,
\quad
y=\widetilde{\calF}_{1}^{{\rm conn}(0)}(\xi(\eta))
=M(\xi(\eta))\frac{\alpha(\bm{\mu})-\beta(\bm{\mu})}{4}\left(\eta-\eta^{-1}\right).
\label{sp_curve_cdt_multi}
\end{align} 
We also define the multi-differentials
\begin{align}
&
\omega_{2}^{(0)}(\eta_{1}, \eta_{2})=B(\eta_{1}, \eta_{2})=
\frac{d\eta_1d\eta_2}{(\eta_1-\eta_2)^2}\,,
\nonumber\\
&
\omega_{N}^{(h)}(\eta_1, \ldots, \eta_N)
=\widetilde{\calF}_{N}^{{\rm conn}(h)}(\xi(\eta_1),\ldots,\xi(\eta_N))\, 
d\xi(\eta_{1}) \cdots d\xi(\eta_{N})
\ \ \textrm{for}\ \ (h,N)\ne (0,2)\,,
\label{multidiff_cdt}
\end{align}
where the bidifferential $B(\eta_{1}, \eta_{2})$ gives the cylinder amplitude \eqref{cyl_cdt} as
\begin{align}
\widetilde{\calF}_{2}^{{\rm conn}(0)}(\xi(\eta_1),\xi(\eta_2))\, d\xi(\eta_1)d\xi(\eta_2)&=
\frac{d\eta_1d\eta_2}{(\eta_1\eta_2-1)^2}
=-B(\eta_1, \eta_2^{-1})
\,.
\label{cyl_local_cdt}
\end{align}
The recursion kernel $K_{\eta_0}(\eta_1)$ in \eqref{top_rec_multi_cdt} is given by
\begin{align}
K_{\eta_0}(\eta_1)&=
\frac{dS_{\eta_0}(\eta_1)}{4\omega_1^{(0)}(\eta_0)}\,,
\\
dS_{\eta_0}(\eta_1)&
=
\int^{\eta_0}_{\eta_0^{-1}} B(\cdot, \eta_1)
=
\frac{\left(\eta_0-\eta_0^{-1}\right) d\eta_1}
{\left(\eta_1-\eta_0\right)\left(\eta_1-\eta_0^{-1}\right)}\,.
\end{align}

For the one-cut disk solution described above, this establishes the correspondence between the SD equations derived from the mode-expanded Hamiltonian of the $m$-th multicritical CDT and the CEO topological recursion.
In Appendices \ref{app:pure_cdt} and \ref{app:m3_cdt}, 
we list several amplitudes computed by the topological recursion \eqref{top_rec_multi_cdt} for the spectral curves \eqref{pure_cdt_disk} of pure CDT and 
\eqref{m3_cdt_disk} of $m=3$ multicritical CDT with $\mu_2=0$, respectively.

\vspace{1cm}
\noindent{\textbf{\Large Acknowledgements}}\\

The authors would like to thank Jan Ambj\o rn and Yasuhiko Yamada for their valuable comments.
We would also like to thank the anonymous referee for helpful comments and suggestions.
This work was supported by JSPS KAKENHI Grant Numbers JP23K22388 and JP25K07278.

\section*{Use of artificial intelligence tools}

During the preparation of this work, the authors used ChatGPT (OpenAI)
for text editing and language polishing. The authors reviewed and edited
all AI-assisted output and take full responsibility for the content of
the publication. ChatGPT was not used for mathematical derivations or
calculations.

\renewcommand{\theequation}{\thesection.\arabic{equation}}
\renewcommand{\thefigure}{\thesection.\arabic{figure}}
\appendix

\section{Proof of Proposition \ref{prop:sd_multi_dt}}
\label{app:proof_prop}

The aim of this appendix is to derive equation \eqref{sp_sd_m_crit_dt_rev} from equation \eqref{sp_sd_m_crit_dt}.

For  $\NN=1$, equation \eqref{sp_sd_m_crit_dt} takes the form
\begin{align}
0&=
\tilde{f}_{2}(\xi, \xi)
+ 2 \left(
\Omega_1(\xi)\, \tilde{f}_{1}(\xi)-
\left(\Omega_1(\xi)\, \tilde{f}_{1}(\xi)\right)_{\mathrm{reg}(\xi)}
\right)
+\frac18 \left(\GG  \xi^{-2}+ 8 \mu_m^2 \xi^{-1}\right)
+C_{0}
\,,
\label{sp_sd_m_crit_dt_N1}
\end{align}
which yields equation \eqref{sd_m_crit_dt_1}.

For $\NN=2$, equation \eqref{sp_sd_m_crit_dt} takes the form
\begin{align}
0&=
\tilde{f}_{3}(\xi_1, \xi_1, \xi_2)
+ 2 \left(
\Omega_1(\xi_1)\, \tilde{f}_{2}(\xi_1, \xi_2)-
\left(\Omega_1(\xi_1)\,  \tilde{f}_{2}(\xi_1, \xi_2)\right)_{\mathrm{reg}(\xi_1)}
\right)
\nonumber\\
&\ \
+2\GG \pder{\xi_2}
\frac{\xi_1^{-1/2}\xi_2^{1/2}
\left(\tilde{f}_{1}(\xi_1)-(-1)^m \mu_m \xi_1^{-1/2}\right)
- \left(\tilde{f}_{1}(\xi_2)-(-1)^m \mu_m \xi_2^{-1/2}\right)}
{\xi_1-\xi_2}
\nonumber\\
&\ \
+\frac18 \left(\GG  \xi_1^{-2}+ 8 \mu_m^2 \xi_1^{-1}\right)
\tilde{f}_{1}(\xi_2)
+C_{1}(\xi_2)
\nonumber\\
&=
\tilde{f}_{3}^{{\rm conn}}(\xi_1, \xi_1, \xi_2)
+
2 \left(\tilde{f}_{1}(\xi_1)+\Omega_1(\xi_1)\right)
\tilde{f}_{2}^{{\rm conn}}(\xi_1, \xi_2)
-2 \left(\Omega_1(\xi_1)\,  \tilde{f}_{2}^{{\rm conn}}(\xi_1, \xi_2)\right)_{\mathrm{reg}(\xi_1)}
\nonumber\\
&\ \
+2\GG \pder{\xi_2}
\frac{\xi_1^{-1/2}\xi_2^{1/2}
\left(\tilde{f}_{1}(\xi_1)+\Omega_1(\xi_1)\right)
- \left(\tilde{f}_{1}(\xi_2)+\Omega_1(\xi_2)\right)}
{\xi_1-\xi_2}
\nonumber\\
&\ \
-2\GG \pder{\xi_2}
\frac{\left(\xi_1^{-1/2}\xi_2^{1/2}
\Omega_1(\xi_1)\right)_{\mathrm{reg}(\xi_1)}
- \Omega_1(\xi_2)_{\mathrm{reg}(\xi_2)}}
{\xi_1-\xi_2}
+\tilde{C}_{1}(\xi_2)\,,
\label{sp_sd_m_crit_dt_N2}
\end{align}
where in the second equality we have used equation \eqref{sp_sd_m_crit_dt_N1}, 
and $\tilde{C}_{1}(\xi_2)$ is a function of $\xi_2$.
Applying the formula
\begin{align}
\pder{\xi_2}
\frac{\xi_1^{-1/2}\xi_2^{1/2} - 1}{\xi_1-\xi_2}
=
\frac{1}{2\sqrt{\xi_1\xi_2}\left(\sqrt{\xi_1}+\sqrt{\xi_2}\right)^2}
=
\Omega_2(\xi_1,\xi_2)\,,
\label{omega2_formula1}
\end{align}
to equation \eqref{sp_sd_m_crit_dt_N2}, 
we obtain equation \eqref{sd_m_crit_dt_2}.

For $\NN=3$, equation \eqref{sp_sd_m_crit_dt} takes the form
\begin{align}
0&=
\tilde{f}_{4}(\xi_1, \bm{\xi}_I)
+ 2 \left(\Omega(\xi_1) \tilde{f}_{3}(\bm{\xi}_I)-
\left(\Omega(\xi_1) \tilde{f}_{3}(\bm{\xi}_I)\right)_{\mathrm{reg}(\xi_1)}
\right)
\nonumber\\
&\ \
+2\GG \sum_{i=2,3} \pder{\xi_i}
\frac{\xi_1^{-1/2}\xi_i^{1/2}\, \tilde{f}_{2}(\bm{\xi}_{I \backslash \{i\}})
- \tilde{f}_{2}(\xi_2, \xi_3)}
{\xi_1-\xi_i}
\nonumber\\
&\ \
+\frac18 \left(\GG  \xi_1^{-2}+ 8 \mu_m^2 \xi_1^{-1}\right)
\tilde{f}_{2}(\xi_2, \xi_3)
-(-1)^m \mu_m \GG \sum_{i=2,3} 
\xi_1^{-1}\xi_i^{-3/2}\,
\tilde{f}_{1}(\bm{\xi}_{I \backslash \{1, i\}})
\nonumber\\
&\ \
+\frac12 \GG^2 
\xi_1^{-1}\xi_2^{-3/2}\xi_3^{-3/2}
+C_{2}(\xi_2, \xi_3)
\nonumber\\
&=
\tilde{f}_{4}^{{\rm conn}}(\xi_1, \bm{\xi}_I)
+2\tilde{f}_{3}^{{\rm conn}}(\bm{\xi}_I)\, \tilde{f}_{1}(\xi_1)
+2\tilde{f}_{2}^{{\rm conn}}(\xi_1, \xi_2)\, \tilde{f}_{2}^{{\rm conn}}(\xi_1, \xi_3)
\nonumber\\
&\ \
+ 2 \left(\Omega(\xi_1) \tilde{f}_{3}^{{\rm conn}}(\bm{\xi}_I)-
\left(\Omega(\xi_1) \tilde{f}_{3}^{{\rm conn}}(\bm{\xi}_I)\right)_{\mathrm{reg}(\xi_1)}
\right)
\nonumber\\
&\ \
+2\GG \sum_{i=2,3} \pder{\xi_i}
\frac{\xi_1^{-1/2}\xi_i^{1/2}\, \tilde{f}_{2}^{{\rm conn}}(\bm{\xi}_{I \backslash \{i\}})
- \tilde{f}_{2}^{{\rm conn}}(\xi_2, \xi_3)}
{\xi_1-\xi_i}
\nonumber\\
&\ \
+\frac12 \GG^2 
\xi_1^{-1}\xi_2^{-3/2}\xi_3^{-3/2}
+\tilde{C}_{2}(\xi_2, \xi_3)\,,
\label{sp_sd_m_crit_dt_N3}
\end{align}
where $I=\{1, 2, 3\}$. In the second equality we used equations \eqref{sp_sd_m_crit_dt_N1} and \eqref{sp_sd_m_crit_dt_N2}, 
and $\tilde{C}_{2}(\xi_2, \xi_3)$ is a function of $\xi_2$ and $\xi_3$.
By applying formulas \eqref{omega2_formula1} and
\begin{align}
\frac14 \xi_1^{-1}\xi_2^{-3/2}\xi_3^{-3/2}
=\Omega_2(\xi_1, \xi_2)\, \Omega_2(\xi_1, \xi_3) +
\sum_{i=2,3}\pder{\xi_i}
\frac{
\Omega_2(\bm{\xi}_{I \backslash\{i\}})-\Omega_2(\xi_2, \xi_3)}{\xi_1-\xi_i}\,,
\label{omega2_formula2}
\end{align}
to equation \eqref{sp_sd_m_crit_dt_N3}, we obtain equation \eqref{sp_sd_m_crit_dt_rev} for $N=3$.
For $\NN \ge 4$, 
equation \eqref{sp_sd_m_crit_dt_rev} can be derived in the same way, or by induction, and 
Proposition \ref{prop:sd_multi_dt} follows.

\section{Proof of Proposition \ref{prop:sd_multi_cdt}}
\label{app:proof_prop_cdt}

The aim of this appendix is to derive equation \eqref{sp_sd_m_crit_cdt_rev} from equation \eqref{sp_sd_multi_cdt}.

For $\NN=1$, equation \eqref{sp_sd_multi_cdt} takes the form
\begin{align}
0&=
\tilde{f}_{2}(\xi, \xi)
+ 
2 \left(
\Omega_1(\xi)\, \tilde{f}_{1}(\xi)-
\left(\Omega_1(\xi)\,  \tilde{f}_{1}(\xi)\right)_{\mathrm{reg}(\xi)}
\right)
+ \xi^{-2}-(-1)^{m}\frac{\mu_m}{g} \xi^{-1}
+ C_{0}
\,,
\label{sp_sd_multi_cdt_N1}
\end{align}
which yields equation \eqref{sp_sd_m_crit_cdt_1}.

For $\NN=2$, equation \eqref{sp_sd_multi_cdt} takes the form
\begin{align}
0&=
\tilde{f}_{3}(\xi_1, \xi_1, \xi_2)
+ 
2 \left(
\Omega_1(\xi_1)\, \tilde{f}_{2}(\xi_1, \xi_2)-
\left(\Omega_1(\xi_1)\,  \tilde{f}_{2}(\xi_1, \xi_2)\right)_{\mathrm{reg}(\xi_1)}
\right)
\nonumber\\
&\ \
+\GG \pder{\xi_2}
\frac{\left(\tilde{f}_{1}(\xi_1)+\xi_1^{-1}\right)
-\left(\tilde{f}_{1}(\xi_2)+\xi_2^{-1}\right)}
{\xi_1-\xi_2}
+ \left(\xi_1^{-2}-(-1)^{m}\frac{\mu_m}{g} \xi_1^{-1}\right)
\tilde{f}_{1}(\xi_2)
+ C_{1}(\xi_2)
\nonumber\\
&=
\tilde{f}_{3}^{{\rm conn}}(\xi_1, \xi_1, \xi_2)
+
2 \left(\tilde{f}_{1}(\xi_1)+\Omega_1(\xi_1)\right)
\tilde{f}_{2}^{{\rm conn}}(\xi_1, \xi_2)
-2 \left(\Omega_1(\xi_1)\,  \tilde{f}_{2}^{{\rm conn}}(\xi_1, \xi_2)\right)_{\mathrm{reg}(\xi_1)}
\nonumber\\
&\ \
+ \GG \pder{\xi_2}
\frac{
\left(\tilde{f}_{1}(\xi_1)+\Omega_1(\xi_1)\right)
- \left(\tilde{f}_{1}(\xi_2)+\Omega_1(\xi_2)\right)}
{\xi_1-\xi_2}
- \GG \pder{\xi_2}
\frac{\Omega_1(\xi_1)_{\mathrm{reg}(\xi_1)}
- \Omega_1(\xi_2)_{\mathrm{reg}(\xi_2)}}
{\xi_1-\xi_2}
\nonumber\\
&\ \
+\tilde{C}_{1}(\xi_2)\,,
\label{sp_sd_multi_cdt_N2}
\end{align}
where in the second equality we used equation \eqref{sp_sd_multi_cdt_N1}, 
and $\tilde{C}_{1}(\xi_2)$ is a function of $\xi_2$.
This gives equation \eqref{sp_sd_m_crit_cdt_2}.
For $\NN \ge 3$, the same argument yields
\eqref{sp_sd_m_crit_cdt_rev}, completing the proof of
Proposition~\ref{prop:sd_multi_cdt}.

\section{List of Amplitudes}
\label{app:list_amp}

\subsection{Pure DT}
\label{app:pure_dt}

For the spectral curve given by \eqref{m2_dt_sp} of pure DT, 
the topological recursion \eqref{top_rec_multi} provides 
the amplitudes as follows:
\begingroup\makeatletter\def\f@size{10}\check@mathfonts
\def\maketag@@@#1{\hbox{\m@th\normalsize\normalfont#1}}%
\begin{align}
&
\widetilde{\calF}_{1}^{{\tinyspace}{\rm conn}(1)}{\negdbltinyspace}(\xi)
=
\frac{2 \xi +5 \sqrt{\cc}}{72 \cc \left(\xi +\sqrt{\cc}\right)^{{5}/{2}}}\,,
\\
&
\widetilde{\calF}_{3}^{{\tinyspace}{\rm conn}(0)}{\negdbltinyspace}(\xi_1,\xi_2,\xi_3)
=
\frac{1}{6\sqrt{\cc} 
\left(\xi_{1}+\sqrt{\cc}\right)^{{3}/{2}}
\left(\xi_{2}+\sqrt{\cc}\right)^{{3}/{2}} \left(\xi_{3}+\sqrt{\cc}\right)^{{3}/{2}}}\,,
\\
&
\widetilde{\calF}_{2}^{{\tinyspace}{\rm conn}(1)}{\negdbltinyspace}(\xi_1,\xi_2)
=
\frac{71 \cc^{2}+91 \left(\xi_{1}+\xi_{2}\right) \cc^{3/2}+\left(35 \xi_{1}^{2}+89 \xi_{1} \xi_{2}+35 \xi_{2}^{2}\right) \cc+28 \xi_{1} \xi_{2} \left(\xi_{1}+\xi_{2}\right) \cc^{1/2}+8 \xi_{1}^{2} \xi_{2}^{2}}
{216\cc^{2}\left(\xi_{1}+\sqrt{\cc}\right)^{{7}/{2}} \left(\xi_{2}+\sqrt{\cc}\right)^{{7}/{2}}}\,,
\\
&
\widetilde{\calF}_{1}^{{\tinyspace}{\rm conn}(2)}{\negdbltinyspace}(\xi)
=
\frac{7 \left(613 \cc^{2}+1006 \cc^{3/2} \xi +816 \cc \xi^{2}+352 \cc^{1/2} \xi^{3}+64 \xi^{4}\right)}
{15552 \cc^{7/2} \left(\xi +\sqrt{\cc}\right)^{{11}/{2}}}\,,
\\
&
\widetilde{\calF}_{1}^{{\tinyspace}{\rm conn}(3)}{\negdbltinyspace}(\xi)
=
\frac{7}{5038848\cc^{6} \left(\xi + \sqrt{\cc}\right)^{{17}/{2}}}
\left(3705145 \cc^{7/2}+11215906 \cc^{3} \xi +17949936 \cc^{5/2} \xi^{2}\right.
\nonumber\\
&\hspace{4.5em}
\left.
+18590240 \cc^{2} \xi^{3}+12875840 \cc^{3/2} \xi^{4}+5779200 \cc \xi^{5}+1523200 \cc^{1/2} \xi^{6}+179200 \xi^{7}\right)\,,
\end{align}
\endgroup
where these amplitudes are also listed in Appendix B.3 of \cite{FMW2025a}.

\subsection{$m=3$ multicritical DT}
\label{app:m3_dt}

For the spectral curve given by \eqref{m3_dt_sp} of $m=3$ multicritical DT with 
specialized cosmological constants, 
the topological recursion \eqref{top_rec_multi} provides 
the amplitudes as follows:
\begingroup\makeatletter\def\f@size{10}\check@mathfonts
\def\maketag@@@#1{\hbox{\m@th\normalsize\normalfont#1}}%
\begin{align}
&
\widetilde{\calF}_{1}^{{\tinyspace}{\rm conn}(1)}{\negdbltinyspace}(\xi)
=-\frac{2 \xi +3 \sqrt{\cc}}{20 \cc^{3/2} \left(\xi + \sqrt{\cc}\right)^{{5}/{2}}}\,,
\\
&
\widetilde{\calF}_{3}^{{\tinyspace}{\rm conn}(0)}{\negdbltinyspace}(\xi_1,\xi_2,\xi_3)
=
\frac{(-1)}{5 \cc \left(\xi_{1}+\sqrt{\cc}\right)^{{3}/{2}} \left(\xi_{2}+\sqrt{\cc}\right)^{{3}/{2}} \left(\xi_{3}+\sqrt{\cc}\right)^{{3}/{2}}}\,,
\\
&
\widetilde{\calF}_{2}^{{\tinyspace}{\rm conn}(1)}{\negdbltinyspace}(\xi_1,\xi_2)
=
\frac{57 \cc^{2}+89 \left(\xi_{1}+\xi_{2}\right) \cc^{3/2}+\left(37 \xi_{1}^{2}+131 \xi_{1} \xi_{2}+37 \xi_{2}^{2}\right) \cc+52 \xi_{1} \xi_{2} \left(\xi_{1}+\xi_{2}\right) \cc^{1/2}+20 \xi_{1}^{2} \xi_{2}^{2}}
{50\cc^{3} \left(\xi_{1}+\sqrt{\cc}\right)^{{7}/{2}}
\left(\xi_{2}+\sqrt{\cc}\right)^{{7}/{2}}}\,,
\\
&
\widetilde{\calF}_{1}^{{\tinyspace}{\rm conn}(2)}{\negdbltinyspace}(\xi)
=
-
\frac{25335 \cc^{2}+76934 \cc^{3/2} \xi +92492 \cc \xi^{2}+51392 \cc^{1/2} \xi^{3}+11024 \xi^{4}}{5000\cc^{5} \left(\xi +\sqrt{\cc}\right)^{{11}/{2}}}\,,
\\
&
\widetilde{\calF}_{1}^{{\tinyspace}{\rm conn}(3)}{\negdbltinyspace}(\xi)
=
\frac{(-11)}{25000\cc^{17/2} \left(\xi +\sqrt{\cc}\right)^{{17}/{2}}}
\left(
2342153 \cc^{7/2}+13475630 \cc^{3} \xi +34197944 \cc^{5/2} \xi^{2}
\right.
\nonumber\\
&\hspace{4.5em}
\left.
+49298752 \cc^{2} \xi^{3}+43395920 \cc^{3/2} \xi^{4}+23246208 \cc \xi^{5}+6998592 \cc^{1/2} \xi^{6}+911744 \xi^{7}\right).
\end{align}
\endgroup

\subsection{Pure CDT}
\label{app:pure_cdt}

The disk and cylinder amplitudes of pure CDT are given in 
\eqref{pure_cdt_disk} and \eqref{pure_cdt_cyl}, respectively, as follows:
\begingroup\makeatletter\def\f@size{10}\check@mathfonts
\def\maketag@@@#1{\hbox{\m@th\normalsize\normalfont#1}}%
\begin{align}
&
\widetilde{\calF}_{1}^{{\rm conn}(0)}(\xi)=
\frac{\xi^2 -\mu_2}{2g} 
+ \frac{1}{\xi + \sqrt{\mu_2}}
+ \frac{\xi+3\sqrt{\mu_2}}{4\mu_2 (\xi + \sqrt{\mu_2})^3}\, g
+ \frac{\xi^3 + 5\sqrt{\mu_2} \xi^2 + 11 \mu_2 \xi +11\mu_2^{3/2}}{8\mu_2^{5/2} (\xi + \sqrt{\mu_2})^5}\, g^2
\nonumber\\
&\hspace{4.5em}
+ \frac{2\xi^5 + 14 \sqrt{\mu_2} \xi^4 + 44\mu_2 \xi^3 + 80 \mu_2^{3/2} \xi^2 + 89 \mu_2^2 \xi + 51 \mu_2^{5/2}}
{16\mu_2^4 (\xi + \sqrt{\mu_2})^7}\, g^3 + O(g^4)
\,,
\\
&
\widetilde{\calF}_{2}^{{\tinyspace}{\rm conn}(0)}{\negdbltinyspace}(\xi_1, \xi_2)
=
\frac{g}{2 \sqrt{\mu_{2}}\, \left(\xi_{1}+\sqrt{\mu_{2}}\right)^{2} \left(\xi_{2}+\sqrt{\mu_{2}}\right)^{2}}
\nonumber\\
&\hspace{4.5em}
+\frac{\left(13 \mu_{2}^{2}+16 \left(\xi_{1}+\xi_{2}\right) \mu_{2}^{{3}/{2}}+2 \left(3 \xi_{1}^{2}+7 \xi_{1} \xi_{2}+3 \xi_{2}^{2}\right) \mu_{2}+4 \xi_{1} \xi_{2} \left(\xi_{1}+\xi_{2}\right) \mu_{2}^{1/2}+\xi_{1}^{2} \xi_{2}^{2}\right)g^2}
{4\mu_{2}^{2} \left(\xi_{1}+\sqrt{\mu_{2}}\right)^{4} \left(\xi_{2}+\sqrt{\mu_{2}}\right)^{4}}
\nonumber\\
&\hspace{4.5em}
+ O(g^3)\,.
\end{align}
\endgroup
Then, the topological recursion \eqref{top_rec_multi_cdt} provides the amplitudes as follows:
\begingroup\makeatletter\def\f@size{10}\check@mathfonts
\def\maketag@@@#1{\hbox{\m@th\normalsize\normalfont#1}}%
\begin{align}
&
\widetilde{\calF}_{1}^{{\tinyspace}{\rm conn}(1)}{\negdbltinyspace}(\xi)
=
\frac{\left(\xi +3 \sqrt{\mu_{2}}\right) \left(\xi^{2}+2 \xi  \sqrt{\mu_{2}}+5 \mu_{2}\right) g^{2}}{32 \mu_{2}^{{5}/{2}} \left(\xi +\sqrt{\mu_{2}}\right)^{5}}
\nonumber\\
&\hspace{4.5em}
+
\frac{\left(7 \xi^{5}+49 \xi^{4} \mu_{2}^{1/2}+154 \xi^{3} \mu_{2}+290 \xi^{2} \mu_{2}^{{3}/{2}}+359 \xi  \mu_{2}^{2}+261 \mu_{2}^{{5}/{2}}\right) g^{3}}{64 \mu_{2}^{4} \left(\xi +\sqrt{\mu_{2}}\right)^{7}}
+ O(g^4)\,,
\\
&
\widetilde{\calF}_{3}^{{\tinyspace}{\rm conn}(0)}{\negdbltinyspace}(\xi_1,\xi_2,\xi_3)
=
\frac{\left(7 \mu_{2}^{{3}/{2}}+5 \mu_{2}\left(\xi_{1}+ \xi_{2}+ \xi_{3}\right)+3 \mu_{2}^{1/2}\left(\xi_{1} \xi_{2}+\xi_{1} \xi_{3}+\xi_{2} \xi_{3}\right)+\xi_{1} \xi_{2} \xi_{3}\right) g^{2}}
{4\mu_{2}^{{3}/{2}} \left(\xi_{1}+\sqrt{\mu_{2}}\right)^{3} \left(\xi_{2}+\sqrt{\mu_{2}}\right)^{3} \left(\xi_{3}+\sqrt{\mu_{2}}\right)^{3}}
+O(g^3)\,,
\\
&
\widetilde{\calF}_{1}^{{\tinyspace}{\rm conn}(2)}{\negdbltinyspace}(\xi)
=
\frac{g^{4}}{2048 \mu_{2}^{{11}/{2}} \left(\xi +\sqrt{\mu_{2}}\right)^{9}}
\left(105 \xi^{7}+945 \xi^{6} \mu_{2}^{1/2}+3885 \xi^{5} \mu_{2}+9765 \xi^{4} \mu_{2}^{{3}/{2}}+16955 \xi^{3} \mu_{2}^{2}\right.
\nonumber\\
&\hspace{13.5em}
\left.
+21555 \xi^{2} \mu_{2}^{{5}/{2}}+19951 \xi  \mu_{2}^{3}+11319 \mu_{2}^{{7}/{2}}\right)
+ O(g^5)\,,
\\
&
\widetilde{\calF}_{1}^{{\tinyspace}{\rm conn}(3)}{\negdbltinyspace}(\xi)
=
\frac{5 g^{6}}{65536 \mu_{2}^{{17}/{2}} \left(\xi +\sqrt{\mu_{2}}\right)^{13}}
\left(5005 \xi^{11}+65065 \xi^{10} \mu_{2}^{1/2}+395395 \xi^{9} \mu_{2}+1496495 \xi^{8} \mu_{2}^{{3}/{2}}\right.
\nonumber\\
&\hspace{13em}
+3970274 \xi^{7} \mu_{2}^{2}+7889882 \xi^{6} \mu_{2}^{{5}/{2}}+12265862 \xi^{5} \mu_{2}^{3}+15360254 \xi^{4} \mu_{2}^{{7}/{2}}
\nonumber\\
&\hspace{13em}
\left.
+15704577 \xi^{3} \mu_{2}^{4}
+12925549 \xi^{2} \mu_{2}^{{9}/{2}}+7963015 \xi  \mu_{2}^{5}+2871011 \mu_{2}^{{11}/{2}}\right)
\nonumber\\
&\hspace{4.5em}
+ O(g^7)\,.
\end{align}
\endgroup
These results are consistent with those of \cite{CDT:SFT:ALWWZ}.

\subsection{$m=3$ multicritical CDT}
\label{app:m3_cdt}

The disk and cylinder amplitudes for the $m=3$ multicritical CDT with $\mu_2=0$ are given in 
\eqref{m3_cdt_disk} and \eqref{m3_cdt_cyl}, respectively, as follows:
\begingroup\makeatletter\def\f@size{10}\check@mathfonts
\def\maketag@@@#1{\hbox{\m@th\normalsize\normalfont#1}}%
\begin{align}
&
\widetilde{\calF}_{1}^{{\rm conn}(0)}(\xi)=
\frac{\xi^{3}+\mu_{3}}{2g} 
+ \frac{1}{\xi+\mu_{3}^{{1}/{3}}}
- \frac{\xi + 2 \mu_{3}^{{1}/{3}}}{3\mu_{3} \left(\xi+\mu_{3}^{{1}/{3}}\right)^{3}}\, g
+\frac{6 \xi^{3}+28 \xi^{2} \mu_{3}^{{1}/{3}}+50 \xi  \mu_{3}^{{2}/{3}}+34 \mu_{3}}
{27\mu_{3}^{{7}/{3}} \left(\xi+\mu_{3}^{{1}/{3}} \right)^{5}}\, g^{2}
\nonumber\\
&\hspace{4.5em}
- \frac{19 \xi^{5}+134 \xi^{4} \mu_{3}^{{1}/{3}}+406 \xi^{3} \mu_{3}^{{2}/{3}}+667 \xi^{2} \mu_{3}+602 \xi  \mu_{3}^{{4}/{3}}+241 \mu_{3}^{{5}/{3}}}
{81\mu_{3}^{{11}/{3}} \left(\xi+\mu_{3}^{{1}/{3}} \right)^{7}}\, g^{3}
+O(g^4)\,,
\\
&
\widetilde{\calF}_{2}^{{\tinyspace}{\rm conn}(0)}{\negdbltinyspace}(\xi_1, \xi_2)
=
-\frac{g}{3 \mu_{3}^{{2}/{3}} \left(\xi_{1}+\mu_{3}^{{1}/{3}}\right)^{2} \left(\xi_{2}+\mu_{3}^{{1}/{3}}\right)^{2}}
\nonumber\\
&\hspace{4.5em}
+\frac{19 \mu_{3}^{{4}/{3}}+26 \left(\xi_{1}+\xi_{2}\right) \mu_{3}+10 \left(\xi_{1}^{2}+3 \xi_{1} \xi_{2}+\xi_{2}^{2}\right) \mu_{3}^{{2}/{3}}+10 \xi_{1} \xi_{2} \left(\xi_{1}+\xi_{2}\right) \mu_{3}^{{1}/{3}}+3 \xi_{1}^{2} \xi_{2}^{2}}
{9\mu_{3}^{2} \left(\xi_{1}+\mu_{3}^{{1}/{3}}\right)^{4} \left(\xi_{2}+\mu_{3}^{{1}/{3}}\right)^{4}} g^{2}
\nonumber\\
&\hspace{4.5em}
+O(g^3)\,.
\end{align}
\endgroup
Then, the topological recursion \eqref{top_rec_multi_cdt} provides the amplitudes as follows:
\begingroup\makeatletter\def\f@size{10}\check@mathfonts
\def\maketag@@@#1{\hbox{\m@th\normalsize\normalfont#1}}%
\begin{align}
&
\widetilde{\calF}_{1}^{{\tinyspace}{\rm conn}(1)}{\negdbltinyspace}(\xi)
=
\frac{\left(\xi^{3}+5 \xi^{2} \mu_{3}^{{1}/{3}}+10 \xi \mu_{3}^{{2}/{3}}+9 \mu_{3}\right) g^{2}}{27 \mu_{3}^{{7}/{3}} \left(\xi+\mu_{3}^{{1}/{3}} \right)^{5}}
\nonumber\\
&\hspace{4.5em}
-
\frac{\left(3 \xi^{5}+23 \xi^{4} \mu_{3}^{{1}/{3}}+77 \xi^{3} \mu_{3}^{{2}/{3}}+144 \xi^{2} \mu_{3}+154 \xi  \mu_{3}^{{4}/{3}}+77 \mu_{3}^{{5}/{3}}\right) g^{3}}{27 \mu_{3}^{{11}/{3}} \left(\xi+\mu_{3}^{{1}/{3}} \right)^{7}}
+ O(g^4)\,,
\\
&
\widetilde{\calF}_{3}^{{\tinyspace}{\rm conn}(0)}{\negdbltinyspace}(\xi_1,\xi_2,\xi_3)
=
\frac{2 \left(4 \mu_{3}+3 \mu_{3}^{{2}/{3}}\left(\xi_{1}+ \xi_{2}+ \xi_{3}\right)+2 \mu_{3}^{{1}/{3}}\left(\xi_{1} \xi_{2}+ \xi_{1} \xi_{3}+ \xi_{2} \xi_{3}\right)+\xi_{1} \xi_{2} \xi_{3}\right) g^{2}}
{9 \mu_{3}^{{5}/{3}} 
\left(\xi_{1}+\mu_{3}^{{1}/{3}}\right)^{3}
\left(\xi_{2}+\mu_{3}^{{1}/{3}}\right)^{3}  \left(\xi_{3}+\mu_{3}^{{1}/{3}}\right)^{3}}
+O(g^3)\,,
\\
&
\widetilde{\calF}_{1}^{{\tinyspace}{\rm conn}(2)}{\negdbltinyspace}(\xi)
=
\frac{\left(55 \xi^{6}+495 \xi^{5} \mu_{3}^{{1}/{3}}+1980 \xi^{4} \mu_{3}^{{2}/{3}}+4565 \xi^{3} \mu_{3}+6498 \xi^{2} \mu_{3}^{{4}/{3}}+5517 \xi  \mu_{3}^{{5}/{3}}+2233 \mu_{3}^{2}\right) g^{4}}
{729 \mu_{3}^{{14}/{3}} \left(\xi+\mu_{3}^{{1}/{3}}\right)^{9}}
\nonumber\\
&\hspace{4.5em}
+ O(g^5)\,,
\\
&
\widetilde{\calF}_{1}^{{\tinyspace}{\rm conn}(3)}{\negdbltinyspace}(\xi)
=
-\frac{5g^{6}}{59049 \mu_{3}^{{23}/{3}} \left(\xi+\mu_{3}^{{1}/{3}} \right)^{13}}
\left(6545 \xi^{11}+78169 \xi^{10} \mu_{3}^{{1}/{3}}+420602 \xi^{9} \mu_{3}^{{2}/{3}}+1325877 \xi^{8} \mu_{3}\right.
\nonumber\\
&\hspace{13em}
\left.
+2623530 \xi^{7} \mu_{3}^{{4}/{3}}+3057873 \xi^{6} \mu_{3}^{{5}/{3}}+996135 \xi^{5} \mu_{3}^{2}-3363360 \xi^{4} \mu_{3}^{{7}/{3}}\right.
\nonumber\\
&\hspace{13em}
\left.
-7087080 \xi^{3} \mu_{3}^{{8}/{3}}-7203823 \xi^{2} \mu_{3}^{3}-4154909 \xi  \mu_{3}^{{10}/{3}}-1113970 \mu_{3}^{{11}/{3}}\right)
\nonumber\\
&\hspace{4.5em}
+ O(g^7)\,.
\end{align}
\endgroup


\begin{thebibliography}{99}

\bibitem{MM:Kaza}
  V.~A.~Kazakov, 
  \emph{The appearance of matter fields
        from quantum fluctuations of 2D-gravity}, 
  Mod. Phys. Lett. A \textbf{4} (1989), 2125.

\bibitem{SFT:GK}
  S.~S.~Gubser and I.~R.~Klebanov, 
  \emph{Scaling functions for baby universes
        in two-dimensional quantum gravity},
  Nucl.\ Phys.\ B \textbf{416}, 827 (1994).

\bibitem{SFT:Watabiki}
  Y.~Watabiki, 
  \emph{Construction of noncritical string field theory
        by transfer matrix formalism in dynamical triangulation},
  Nucl.\ Phys.\ B \textbf{441} (1995), 119,
  [arXiv:hep-th/9401096 [hep-th]].

\bibitem{SFT:AW}
  J.~Ambj{\o}rn and Y.~Watabiki, 
  \emph{Noncritical string field theory for 2d quantum gravity
        coupled to (p,q)-conformal fields},
  Int. J. Mod. Phys. A \textbf{12} (1997), 4257,
  [arXiv:hep-th/9604067 [hep-th]].

\bibitem{MM:FKN}
  M.~Fukuma, H.~Kawai and R.~Nakayama,
  \emph{Continuum Schwinger-Dyson Equations and Universal Structures
        in Two-dimensional Quantum Gravity},
  Int. J. Mod. Phys. A \textbf{6} (1991), 1385-1406,
  \emph{Infinite dimensional Grassmannian structure of
        two-dimensional quantum gravity},
  Commun. Math. Phys. \textbf{143} (1992), 371-404.

\bibitem{MM:DVV}
R.~Dijkgraaf, H.~L.~Verlinde and E.~P.~Verlinde,
  \emph{Loop equations and Virasoro constraints 
        in nonperturbative 2-D quantum gravity},
  Nucl. Phys. B \textbf{348} (1991), 435-456.

\bibitem{FMW2025a}
H.~Fuji, M.~Manabe and Y.~Watabiki,
  \emph{Dynamical Triangulations for 2D Pure Gravity and Topological Recursion},
JHEP  \textbf{05} (2026), 208
[arXiv:2509.18916 [hep-th]].

\bibitem{Eynard:2004mh}
B.~Eynard,
\emph{Topological expansion for the 1-Hermitian matrix model correlation functions},
JHEP \textbf{11}, 031 (2004)
[arXiv:hep-th/0407261 [hep-th]].

\bibitem{Chekhov:2006vd}
L.~Chekhov, B.~Eynard and N.~Orantin,
\emph{Free energy topological expansion for the 2-matrix model},
JHEP \textbf{12} (2006), 053,
[arXiv:math-ph/0603003 [math-ph]].

\bibitem{Eynard:2007kz}
B.~Eynard and N.~Orantin,
\emph{Invariants of algebraic curves and topological expansion},
Commun. Num. Theor. Phys. \textbf{1} (2007), 347-452,
[arXiv:math-ph/0702045 [math-ph]].

\bibitem{CDT:AL}
  J.~Ambj\o rn, R.~Loll, 
  \emph{Non-perturbative Lorentzian Quantum Gravity,
Causality and Topology Change},
  Nucl.\ Phys.\ B \textbf{536} (1998) 407-434, [hep-th/9805108].

\bibitem{CDT:SFT:ALWWZ}
  J.~Ambj\o rn, R.~Loll, Y.~Watabiki, W.~Westra, S.~Zohren, 
  \emph{A String Field Theory based on Causal Dynamical Triangulations},
  JHEP 05 (2008) 032, [arXiv:0802.0719 [hep-th]]; 
  \emph{A Matrix Model for 2D Quantum Gravity defined by 
        Causal Dynamical Triangulations},
  Phys.\ Lett.\ B \textbf{665} (2008), 252-256, [arXiv:0804.0252 [hep-th]].

\bibitem{CDT:MM:ALWWZ}
  J.~Ambj\o rn, R.~Loll, Y.~Watabiki, W.~Westra, S.~Zohren, 
  \emph{A new continuum limit of matrix models},
  Phys.\ Lett.\ B \textbf{670} (2008), 224-230, [arXiv:0810.2408 [hep-th]].

\bibitem{CDT:MM:ABW}
J.~Ambj\o rn, T.~Budd, Y.~Watabiki,
  \emph{Scale-dependent Hausdorff dimensions in 2d gravity},
  Phys.\ Lett.\ B \textbf{736} (2014), 339-343, [arXiv:1406.6251].

\bibitem{JM83}
M.~Jimbo and T.~Miwa, 
\emph{Solitons and Infinite Dimensional Lie Algebras}, 
Publ. Res. Inst. Math. Sci. \textbf{19} (1983), no. 3, pp. 943-1001.

\bibitem{CDT:SFT:AW}
  J.~Ambj\o rn and Y.~Watabiki,
  \emph{A model for emergence of space and time},
  Phys.\ Lett.\ B {\bf 749} (2015) 149
  [arXiv:1505.04353 [hep-th]].

\bibitem{Atkin:2012ka}
M.~R.~Atkin and S.~Zohren,
\emph{On the Quantum Geometry of Multi-critical CDT},
JHEP \textbf{11} (2012), 037
[arXiv:1203.5034 [hep-th]].

\bibitem{CDT:AGGS}
  J.~Ambj{\o}rn, L.~Glaser, A.~G\"{o}rlich, Y.~Sato,
  \emph{New multicritical matrix models and multicritical 2d CDT},
  Phys.\ Lett.\ B {\bf 712} (2012) 109
  [arXiv:1202.4435v2 [hep-th]].

\bibitem{FMW3}
H.~Fuji, M.~Manabe and Y.~Watabiki,
\emph{A Hamiltonian Formalism for Topological Recursion},
JHEP  \textbf{08} (2026), 81
[arXiv:2512.14059 [math-ph]].

\bibitem{MM:Stau}
  M.~Staudacher, 
  \emph{The Yang-Lee Edge Singularity on a Dynamical Planar Random Surface},
  Nucl.\ Phys.\ B \textbf{336} (1990), 349.

\bibitem{Ishibashi:1993nq}
N.~Ishibashi and H.~Kawai,
\emph{String field theory of noncritical strings},
Phys. Lett. B \textbf{314} (1993), 190-196
[arXiv:hep-th/9307045 [hep-th]].

\bibitem{Gregori:2021tvs}
P.~Gregori and R.~Schiappa,
\emph{From minimal strings towards Jackiw{\textendash}Teitelboim gravity: on their resurgence, resonance, and black holes},
Class. Quant. Grav. \textbf{41} (2024) no.~11, 115001
[arXiv:2108.11409 [hep-th]].

\bibitem{Seiberg:2003nm}
N.~Seiberg and D.~Shih,
\emph{Branes, rings and matrix models in minimal (super)string theory},
JHEP \textbf{02} (2004), 021
[arXiv:hep-th/0312170 [hep-th]].

\bibitem{Moore:1991ir}
G.~W.~Moore, N.~Seiberg and M.~Staudacher,
\emph{From loops to states in 2-D quantum gravity},
Nucl. Phys. B \textbf{362} (1991), 665-709.

\bibitem{Brini:2010fc}
A.~Brini, M.~Marino and S.~Stevan,
\emph{The Uses of the refined matrix model recursion},
J. Math. Phys. \textbf{52} (2011), 052305,
[arXiv:1010.1210 [hep-th]].

\bibitem{Fuji:2023wcx}
H.~Fuji and M.~Manabe,
\emph{Some Generalizations of Mirzakhani's Recursion and Masur-Veech Volumes via Topological Recursions},
SIGMA \textbf{20} (2024), 043
[arXiv:2303.14154 [math-ph]].

\bibitem{Stanford:2017thb}
D.~Stanford and E.~Witten,
\emph{Fermionic Localization of the Schwarzian Theory},
JHEP \textbf{10} (2017), 008
[arXiv:1703.04612 [hep-th]].

\bibitem{Saad:2019lba}
P.~Saad, S.~H.~Shenker and D.~Stanford,
\emph{JT gravity as a matrix integral},
[arXiv:1903.11115 [hep-th]].

\bibitem{Eynard:2007fi}
B.~Eynard and N.~Orantin,
\emph{Weil-Petersson volume of moduli spaces, Mirzakhani's recursion and matrix models},
[arXiv:0705.3600 [math-ph]].

\bibitem{Mirzakhani:2006fta}
M.~Mirzakhani,
\emph{Simple geodesics and Weil-Petersson volumes of moduli spaces of bordered Riemann surfaces},
Invent. Math. \textbf{167} (2006) no.1, 179-222.



\end{thebibliography}
\end{document}